\documentclass[twocolumn,pra,aps,superscriptaddress,showpacs]{revtex4-1}

\usepackage{color}
\usepackage[english]{babel}

\usepackage{amsmath,amssymb,amsfonts}

\usepackage{graphicx}

\definecolor{myblue}{rgb}{0.2,0.2,0.8}
\definecolor{myblack}{rgb}{0,0,0}

\usepackage[colorlinks=true,citecolor=myblue,linkcolor=myblack]{hyperref}

 \newcommand{\ket}[1]{|#1\rangle}
 \newcommand{\bra}[1]{\langle #1|}
 \newcommand{\braket}[2]{\langle #1|#2\rangle}
 
\def\oned{\mathrm{1d}}

\newcommand{\src}{\mathrm{(s)}}
\newcommand{\trg}{\mathrm{(t)}}
\newcommand{\dtc}{\mathrm{(d)}}
\newcommand\Tr{\mathrm{Tr}}

\def\dd{\mathord{\rm d}}
\def\ee{\mathord{\rm e}}
\def\ii{\mathord{\rm i}}
\newcommand{\rs}{\mathrm{s}}
\newcommand{\rd}{\mathrm{d}}
\newcommand{\rg}{\mathrm{g}}

\newcommand{\rt}{\mathrm{t}}
\renewcommand{\aa}{\mathrm{a}}
\newcommand{\bb}{\mathrm{b}}

\begin{document}

\title{Generation of single and two-mode multiphoton states in waveguide QED.}

\author{V. Paulisch}
\affiliation{Max-Planck-Institut f\"{u}r Quantenoptik, Hans-Kopfermann-Str. 1, 85748 Garching, Germany }

\author{H. J. Kimble}
\affiliation{Max-Planck-Institut f\"{u}r Quantenoptik, Hans-Kopfermann-Str. 1, 85748 Garching, Germany }
\affiliation{Norman Bridge Laboratory of Physics 12-33}
\affiliation{California Institute of Technology, Pasadena, CA 91125, USA}

\author{J. I. Cirac}
\affiliation{Max-Planck-Institut f\"{u}r Quantenoptik, Hans-Kopfermann-Str. 1, 85748 Garching, Germany }

\author{A. Gonz\'{a}lez-Tudela}
\affiliation{Max-Planck-Institut f\"{u}r Quantenoptik, Hans-Kopfermann-Str. 1, 85748 Garching, Germany }

\date{\today}

\begin{abstract}
Single and two-mode multiphoton states are the cornerstone of many quantum technologies, e.g., metrology. In the optical regime these states are generally obtained combining heralded single-photons with linear optics tools and post-selection, leading to inherent low success probabilities. In a recent paper~\cite{gonzaleztudela17a}, we design several protocols that harness the long-range atomic interactions induced in waveguide QED to improve fidelities and protocols of single-mode multiphoton emission. Here, we give full details of these protocols, revisit them to simplify some of their requirements and also extend them to generate two-mode multiphoton states, such as Yurke or NOON states.
\end{abstract}
 
\maketitle

\section{Introduction}\label{sec:intro}

Single-mode photonic states play a capital role in many quantum technologies. For example, single-photons are the prime candidate for the high fidelity exchange of quantum information between distant parties, e.g., in a quantum network \cite{kimble08a}, whereas states with large (and fixed) photon numbers~\cite{caves81,holland93,giovannetti04a} can be used to overcome certain limitations of classical light, e.g., the Standard Quantum Limit of phase sensitivity. Experimentally, the generation of single-photons can be achieved with high-fidelities and efficiencies~\cite{lounis05a} due to impressive progress in the integration of single quantum emitters with cavities or waveguides systems~\cite{xu99a,painter99a,hughes04a,rao07a,laucht12a,arcari14a,somaschi16a,reiserer15}. However, extending these protocols to the multiphoton regime is much more challenging and their efficient generation is still an open question~\cite{dellanno06}. Most advanced methods \cite{wang16a} rely on combining heralded single-photons with beam-splitters 
and post-selection to achieve high-photon numbers. Unfortunately, these protocols suffer from inherent low probabilities which scale exponentially with the photon number, limiting most of their applications. Therefore, the search for alternative protocols is a very timely issue.

Atomic ensembles trapped near nanophotonic waveguides~\cite{vetsch10a,goban13a,beguin14a,goban15a,sorensen16a,corzo16a,solano17} are well suited for the generation of photonic states: 
i) atoms have naturally several ground states which can be used as quantum memories, together with several optically excited states which can be used to trigger the emission to waveguide photons; 
ii) the confinement of guided photons increase the decay rates into waveguide modes, $\Gamma_{\oned}$, compared to free space ones, $\Gamma^*$, which can be further boosted with structured waveguides by designing regions of slow-light~\cite{goban13a,goban15a,laucht12a,lodahl15};
iii) the interaction with one-dimensional guided photons leads to long-range interactions~\cite{kien05a} that in the \emph{atomic mirror} configuration~\cite{chang12a,sorensen16a,corzo16a} provides an enhancement of the decay rates of certain atomic states, e.g.,  symmetric Dicke states, $\ket{\phi^e_m} \propto \mathrm{sym}\{ \ket{e}^{\otimes m} \ket{g}^{\otimes N-m}\}$ where $g$ denotes an atomic ground state and $e$ and an optically excited state.
When excited, these Dicke states decay into their ground state, emitting $m$ waveguide photons, which in the low excitation limit [$m\ll N$], travel as a single-mode in the waveguide \cite{porras08a,gonzaleztudela15a}. The difficulty of generating multiphoton states in this scenario thus translates to generating Dicke states with a given excitation number $m$.

In a recent proposal~\cite{gonzaleztudela17a}, we exploited the atom waveguide characteristics (i-iii) to obtain Dicke states with large (and fixed) photon number with high fidelities and probabilities. The common step of the protocols is the probabilistic \emph{loading} of a single collective excitation into an atomic ensemble by using an auxiliary atom to transfer one excitation through the waveguide, which can be done with probability $p$ when performing a measurement to herald the successful transfer. Then, to achieve high-photon numbers, one can either apply the \emph{loading} protocol $m$ times to add the excitations into the same level, that results in an overall probability $p^m$. The other alternative consists in loading the collective atomic excitations into several hyperfine levels and combining them a posteriori to circumvent the exponential decrease of probabilities~\cite{fiurasek05a}. 

In this paper we will revisit these protocols with a three-fold intention:
i) first, we will give full details on the protocols of Ref.~\cite{gonzaleztudela17a} and numerically certify that the approximations made in Ref.~\cite{gonzaleztudela17a} capture accurately the scalings by solving exactly the corresponding master equation governing the evolution;
ii) we develop a new protocol that simplifies part of the requirements of the previous ones, e.g., the use of a transition with strongly supressed decay rate;
iii) finally, we show how to adapt these protocols to generate directly two-mode multiphoton states, focusing on examples with metrological interest such as NOON~\cite{giovannetti04a} or Yurke states~\cite{yurke86}.

The paper is organized as follows: in Section~\ref{sec:waveguide} we introduce the atom-waveguide setups, explaining the different ingredients that we use to develop our protocols and their theoretical description. In Section~\ref{sec:theory}, we introduce a general framework on how we analyze the figures of merit of the protocols, namely, the errors (or infidelities) and success probabilities. Then, in Section \ref{sec:single}, we revisit the protocols of Ref.~\cite{gonzaleztudela17a}, providing a detailed description and the numerical certification of the results. We also include here a description of a new protocol in Section~\ref{sec:new}, which simplifies some of the requirements of previous ones. Finally, in Section~\ref{sec:twomode} we show how to extend our protocols to generate two-mode states with metrological interest and conclude in Section~\ref{sec:outlook}.

\section{Atom-waveguide QED}\label{sec:waveguide}

The setup that we consider are atomic ensembles close to waveguides in which one (or several) atomic optical transitions are interacting with the guided modes of these structures, or emitting to free-space photons. We also use additional laser or microwave fields to control the state of the atomic ensemble. These three ingredients can be described in a common framework in which the system dynamics, $\rho$, is governed in general by the following Master Equation~\cite{gardiner_book00a} (with $\hbar\equiv 1$ throughout the paper):
\begin{align}\label{eq:rho}
\dot \rho =
	- \ii \left[ H_\mathrm{L} , \rho \right]
	+ \mathcal{L}_\mathrm{coll} \left[ \rho \right]
	+ \mathcal{L}_\mathrm{ind} \left[ \rho \right],
\end{align}
corresponding to the coherent dynamics induced by either lasers/microwave fields, $H_L$, the collective dissipation induced by the waveguide modes, $\mathcal{L}_\mathrm{coll} \left[ \rho \right]$, and the free-space spontaneous emission, $\mathcal{L}_\mathrm{ind} \left[ \rho \right]$. We assume the spontaneous emission into free space to be individual for all atoms in the ensemble, i.e., for an optical transition $g-e$, to read
\begin{equation}\label{eq:lind}
\mathcal{L}_\mathrm{ind} \left[ \rho \right]
	= \frac{\Gamma^*}{2} \sum_n
	\left( 
	2 \sigma_{ge}^n \rho \sigma_{eg}^n - \sigma_{eg}^n \sigma_{ge}^n \rho - \rho \sigma_{eg}^n \sigma_{ge}^n
	\right).
\end{equation}
This assumption may break in the case of subwavelength spacing between the emitters \cite{asenjogarcia17a}, where strongly subradiant states with highly suppressed spontaneous emission emerge. The combination of these states with our protocol represents an exciting perspective to improve our protocols that we leave for further studies.

Our protocols are designed for the \emph{atomic mirror} configuration~\cite{chang12a,corzo16a,sorensen16a}, in which the atomic positions are commensurate with the wavelength mediating the interactions, such that the collective Lindblad term, $\mathcal{L}_\mathrm{ind} \left[ \rho \right]$, for a given optical transition $g-e$ reads
\begin{equation}\label{eq:coll}
\mathcal{L} _\mathrm{coll} \left[ \rho \right]
	= \frac{\Gamma_\oned}{2} 
	\left( 2 S_{ge} \rho S_{eg} - S_{eg} S_{ge} \rho - \rho S_{eg} S_{ge}\right),
\end{equation}
where we have introduced the collective operator $S_{\alpha \beta} = \sum_n \sigma_{\alpha \beta}^n$, where $\sigma_{\alpha \beta}^n = \ket{\alpha}_n \bra{\beta}$ and where the sum runs over all emitters. One figure of merit to characterize these systems is the so-called \emph{Purcell factor}, $P_\oned=\frac{\Gamma_\oned}{\Gamma^{*}}$, which is of the order of $P_\oned \sim 1 - 100$ in engineered dielectrics \cite{hughes09a,laucht12,lodahl15,yu14a,goban15a}. The other important parameter is the number of atoms, $N$, which is of the order of $1000$ [$\sim 3$] in unstructured [structured] waveguides. Along this paper we will work in the limits of $N\gg m$ and $P_\oned\gg 1$, unless stated otherwise. 

The last ingredient, that is, the lasers/microwave fields acting collectively on a given ensemble can be written in full generality as
\begin{equation}
H_\mathrm{L} 
	= \frac{1}{2} \Omega_{\alpha \beta} \left(  S_{\alpha \beta}+ S_{\beta \alpha} \right),
\end{equation}
where $\Omega_{\alpha \beta}$ is the amplitude of the coupling between the states $\alpha$ and $\beta$ of the ensemble.

Eq.~\ref{eq:rho} can be generally rewritten as the sum of a non-hermitian evolution and jump operators, i.e.,
\begin{equation} \label{eq:effrho}
\dot{\rho}
	= -i\left[H_\mathrm{eff},\rho\right]+J_\mathrm{coll}[\rho]+J_\mathrm{ind}[\rho]\,,
\end{equation}
where $H_\mathrm{eff}$ contains the effective (non-hermitian) Hamiltonian with the unitary evolution of $H_L$ and the non-unitary evolution from the Lindblad terms. For example, for the collective and individual dissipation considered in Eqs.~(\ref{eq:lind}) and (\ref{eq:coll}) it is given by
\begin{equation}\label{eq:eff}
H_\mathrm{eff}
	= H_L - i\frac{\Gamma_\oned}{2}S_{eg} S_{ge}-i\frac{\Gamma^*}{2}\sum_{n}\sigma^n_{ee}\,.
\end{equation}

The terms $J_{\mathrm{ind},\mathrm{coll}}[\rho]$ describe the quantum jump evolution and are written as
\begin{align}\label{eq:qjump}
J_\mathrm{coll}[\rho]
	&=\Gamma_\oned S_{ge}\rho S_{eg}\,,\\
J_\mathrm{ind}[\rho]
	&=\Gamma^* \sum_n \sigma^n_{ge}\rho \sigma^n_{eg}\,.
\end{align}

Using this separation, a formal solution of Eq.~(\ref{eq:rho}) can be split into the evolution driven by $H_\mathrm{eff}$, i.e., $S(t,t_0)\rho=e^{-i H_{\mathrm{eff}} t}\rho e^{i H_{\mathrm{eff}}^\dagger t}$, and the one resulting from quantum jumps evolution which reads
\begin{align} \label{eq:unravel}
\rho(t)
	&= S(t,t_0)\rho(t_0) \\
	& +\sum_{n=1}^\infty \int_{t_0}^t \dd t_n \cdots\int_{t_0}^{t_2} \dd t_1 
			S(t,t_n) J \cdots J S(t_1,t_0)\rho(t_0) \,. \nonumber
\end{align}
where $J[\rho]=J_\mathrm{ind}[\rho]+J_\mathrm{coll}[\rho]$. The $n$-th order of the sum corresponds to the evolution where $n$ quantum jumps have occured. As we explain below, in all the cases we consider here, there is at most a single excitation in the system such that the terms with $n>1$ will be identically zero.

\section{Guidelines to analyze the protocols}\label{sec:theory}

\begin{figure}[t]
	\centering
	\includegraphics[width=0.45\textwidth]{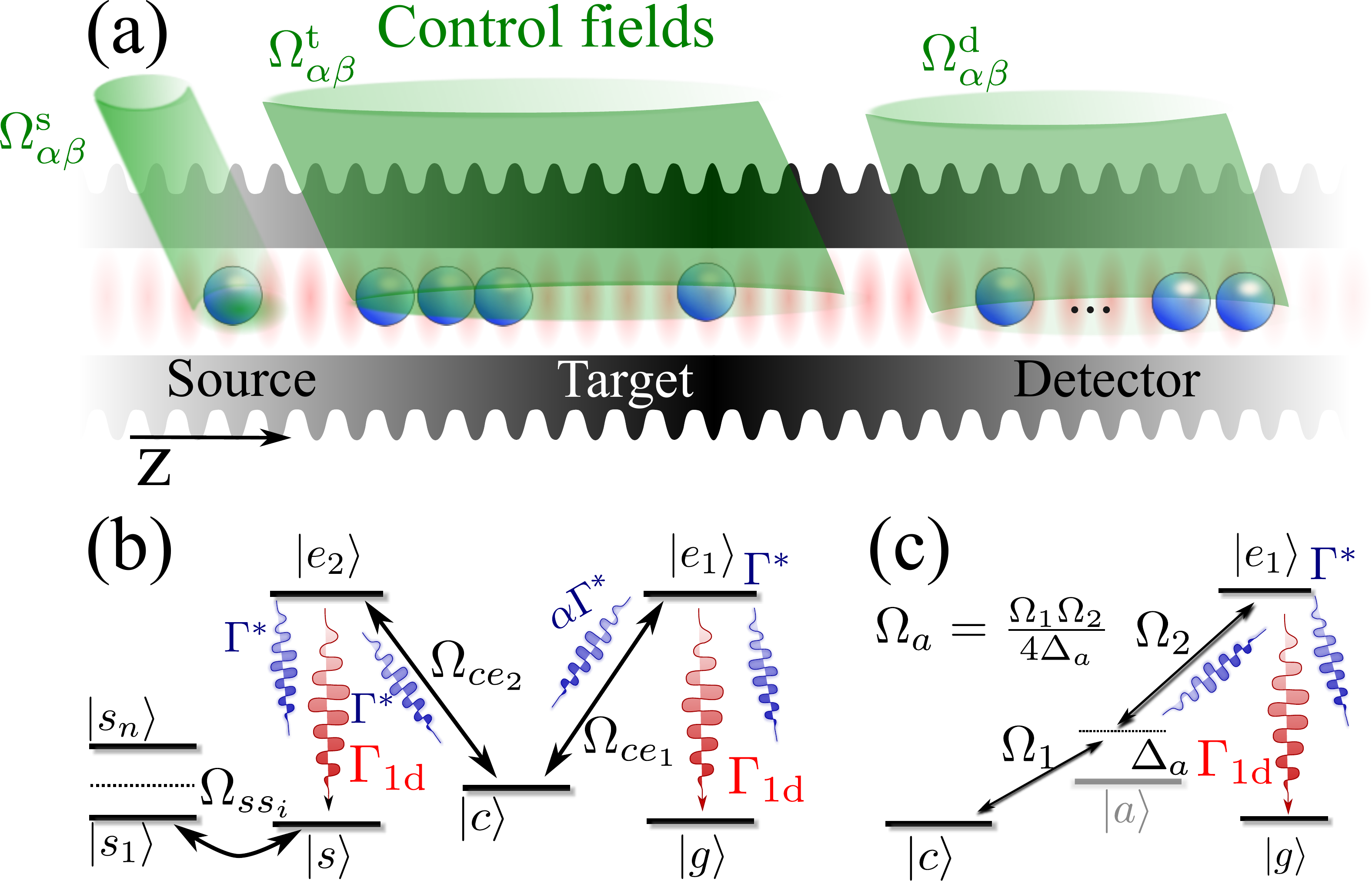}
	\caption{
		(a) General Setup of atoms coupled to a one-dimensional photonic waveguide. We depict the splitting into source, target and detector ensembles with $1$, $N$ and $N_\rd$ emitters respectively, which can be controlled by external fields independently.
		(b) Internal level structure of emitters in which the transitions $g - e_1$ and $s - e_2$ are coupled to a waveguide mode. 
		The Rabi coupling between atomic states $\alpha$ and $\beta$, denoted by $\Omega_{\alpha\beta}$ can be obtained through laser or microwave fields.
		(c) The supression of the decay rate corresponding to one transition can be implemented by using a far off-resonance two-photon transition via a metastable state.
	}
	\label{fig:Scheme}
\end{figure}

The common configuration of all the protocols analyzed in this paper is to have three different sets of atoms as depicted in Fig.~\ref{fig:Scheme}, where each set can be independently addressed. The goal of the protocols is the generation of $m$ collective excitations in an atomic ensemble with $N$ atoms, called the \emph{target} ensemble. We denote such state as $\ket{\phi_m}$. To do it, we will herald the transfer of single collective excitations from the \emph{source} atom one by one, by measuring a given state, $\ket{\psi_{\mathrm{her}}}$, in an auxiliary atomic ensemble with $N_d$ atoms that we call \emph{detector} ensemble.

The first step to analyze the protocols consists in calculating the probability of success and fidelities of heralding the addition of single collective excitation to the \emph{target} ensemble, assuming it already contains $m$ excitations stored in some hyperfine level. This means we initialize in a state, $\ket{\Psi_0}=\ket{e}_\rs \ket{\phi_m}_\rt \ket{\psi_0}_\rd$, where $\ket{\psi_0}_\rd$ is the initial state of the \emph{detector} ensemble. After letting the system evolve for a time $T$ under the interaction induced by the waveguide, lasers, etc., the atomic state is described by the density matrix $\rho(T)=S(T,0)\rho_m(0)$ with $\rho_m(0)=\ket{\Psi_0} \bra{\Psi_0}$. We calculate the probability of succeeding in the heralding,
\begin{equation}
p_{m\rightarrow m+1}
	=\Tr \left[\bra{\psi_{\mathrm{her}}}_\rd S(T,0) \rho_m(0) \ket{\psi_{\mathrm{her}}}_\rd \right]\,,
\end{equation}
and note, that due to the heralding measurement, we only need to consider the non-hermitian evolution and can assume that no jump has occured.

However, the heralded state is in general not perfect and we might have created states in the \emph{target} ensemble other than the desired one, i.e., $\ket{\phi_{m+1}}_\rt$. This is quantified by the fidelity after heralding, which reads
\begin{equation}
F_{m\rightarrow m+1}
	=\frac{\bra{g}_\rs\bra{\phi_{m+1}}_\rt \bra{\psi_{\mathrm{her}}}_\rd 
			S\rho_m(0) \ket{g}_\rs \ket{\phi_{m+1}}_\rt \ket{\psi_{\mathrm{her}}}_\rd }{p}\,,
\end{equation}
whereas the error or infidelity is defined as $I_{m\rightarrow m+1}=1-F_{m\rightarrow m+1}$ for adding a single excitation. Once we have analyzed this general step, we consider the complete process to arrive from $\ket{\phi_0}\rightarrow \ket{\phi_1}\rightarrow\dots\rightarrow \ket{\phi_m}$ and calculate the average number of repetitions, $R_m$ to arrive to $\ket{\phi_m}$ and the average infidelity of the process that we denote as $I_m$. 
Both $R_m$ and $I_m$ will depend on the way we merge the excitations and changes with the different protocols, as explained in the next Sections. As the error in atom-photon mapping of the Dicke states scales as $\epsilon_{ph} \sim \frac{m}{NP_\oned}$ \cite{gonzaleztudela17a}, our goal is to design protocols which follow at least the same scaling, while keeping a high overall probability, or small number of repetitions $R_m$.

\section{Generating single-mode multi-photon states\label{sec:single}}

In this Section we include the discussion of three different protocols to generate $m$ symmetric collective excitations in the ground state, e.g., $s$, of the \emph{target} ensemble, i.e., $\ket{\phi^s_m}_\rt \propto S_{sg}^m \ket{g^{\otimes N}}_\rt$. In Section~\ref{sec:protocol1} and~\ref{sec:protocol2} we discus two protocols that were already introduced in Ref.~\cite{gonzaleztudela17a}. Here, we will provide more details on how to characterize them, and make exact numerical simulations to confirm the analytical scalings provided in the original reference. In Section~\ref{sec:new} we discuss a new protocol that simplifies some of the requirements of the previous ones by using two guided modes.

\subsection{Protocol 1: Excitations added directly with a single guided mode \label{sec:protocol1}}

This protocol consists in \emph{loading} $m$ symmetric collective excitations into the same hyperfine level of the target ensemble, heralding the transfer after each excitation by detecting a change in the \emph{detector} ensemble. The \emph{loading} of a single excitation consists itself of two steps:
i) first, we transfer an excitation between the \emph{source} atom into the \emph{target} ensemble, and
ii) moving the excitation collectively between metastable states of the \emph{target} ensemble while inducing a change in the detector ensemble, on which we can perform a heralding measurement.
In this Section we explain the atom-waveguide resources required, give an overview of the protocol, and then analyze the scaling of the different parts.

\textbf{Atom-waveguide resources}

We require an atomic level structure as depicted in Fig.~\ref{fig:Scheme} (b): two optical transitions $g-e_1$ and $s-e_2$ are coupled to the same waveguide mode, with rate $\Gamma_\oned$. The excited states $e_1,e_2$ can be also coupled to $c$ with a Raman laser $\Omega_{ce_{1/2}}$ respectively. Finally, an extra metastable, $s_1$, state is required which can be coupled to $s$ through a microwave field $\Omega_{s s_1}$.

The excited states, $e_1,e_2$ can decay to the ground states $s,c,g$ with a rate $\Gamma^*$. However, it will be important for the protocol that the spontaneous decay rate of the $c-e_1$ transition is suppressed by a factor $\alpha$  \cite{enk97,porras08a,borregaard15a}, which can be obtained, e.g., by using an off-resonant two-photon transition through an intermediate state as schematically depicted in Fig.~\ref{fig:Scheme}(c).

\textbf{Overview of the protocol}

To minimize errors and, at the same time, obtain a large success probability of the \emph{loading} of single collective excitations, we need to add a few preparatory and intermediate steps beyond the two steps that we described above. Assuming that we start with $\ket{c}_\rs\ket{\phi^s_m}_\rt\ket{s}^{\otimes N_d}$, the full protocol reads:
\begin{enumerate}
	\item [(a)]
	First, we collectively transfer the excitations of the \emph{target/detector} ensemble in $s$ to $s_1$ with $\Omega_{ss_1}$. This is needed to effectively decouple the \emph{detector ensemble} from the dynamics when the excitation is first transfered from the \emph{source} to the \emph{target} ensemble. This results in
	\begin{equation}
	\ket{c}_\rs  \ket{\phi_m^\rs}_\rt \ket{s}^{\otimes N_\rd}_\rd
	\rightarrow
	\ket{c}_\rs \ket{\phi_m^{\rs_1}}_\rt \ket{s_1}^{\otimes N_\rd}_\rd.
	\end{equation}
	
	\item [(b)] Next, the \emph{source} atom is excited by a fast $\pi$-pulse, such that:
	\begin{equation}
	\ket{c}_\rs \ket{\phi_m^{\rs_1}}_\rt \ket{s_1}^{\otimes N_\rd}_\rd
	\rightarrow
	\ket{e_1}_\rs \ket{\phi_m^{\rs_1}}_\rt \ket{s_1}^{\otimes N_\rd}_\rd.
	\end{equation}
	
	\item [(c)] Just after that, the laser $\Omega_{ce_1}^\trg$ in the target ensemble is switched on under the Quantum Zeno conditions which we explain below. As we show below, by choosing the appropriate timing, the excitation can be completely transfered to the \emph{target} ensemble as
	\begin{equation}
	\ket{e_1}_\rs \ket{\phi_m^{\rs_1}}_\rt \ket{s_1}^{\otimes N_\rd}_\rd
	\rightarrow
	\ket{g}_\rs S_{cg}^{(\rt)}\ket{\phi_m^{\rs_1}}_\rt \ket{s_1}^{\otimes N_\rd}_\rd\,,
	\end{equation}
	where the superindex $\trg$ denotes the ensemble on which the operator acts.
	
	\item [(d)] To be able to herald the successful transfer, we need to induce a change in the detector ensemble on which a heralding measurement can be performed. For this purpose, we use the second optical transition $e_2-s$. In order to make the target and detector ensemble interact, we move back the excitations in both the \emph{target/detector} ensemble from $s_1\rightarrow s$ by applying a $\pi$-pulse with $\Omega_{s s_1}$. Furthermore we decouple the source atom by moving $g \rightarrow c$. The resulting state is
	\begin{equation}
	\ket{g}_\rs S_{cg}^{(\rt)}\ket{\phi_m^{\rs_1}}_\rt \ket{s_1}^{\otimes N_\rd}_\rd
	\rightarrow
	\ket{c}_\rs S_{cg}^{(\rt)}\ket{\phi_m^{\rs}}_\rt \ket{s}^{\otimes N_\rd}_\rd.
	\end{equation}
	
	\item [(e)] Next, the target ensemble is excited by a fast $\pi$-pulse on the transition $c - e_2$ and the excitation is transferred to the detector ensemble again under the Quantum Zeno Dynamics with an external driving $\Omega_{c e_2}^\dtc$ on the $c - e_2$ transition of the detector ensemble. This joins the additional excitation to the collective state of the target ensemble, i.e.
	\begin{equation}
	\ket{c}_\rs S_{e_2 g}^{(\rt)}\ket{\phi_m^{\rs}}_\rt \ket{s}^{\otimes N_\rd}_\rd
	\rightarrow
	\ket{c}_\rs \ket{\phi_{m+1}^{\rs}}_\rt S_{cs}^{(\rd)} \ket{s}^{\otimes N_\rd}_\rd.
	\end{equation}
	
	\item [(f)] The final step consists of a measurement on the state $c$ of the detector ensemble. There can only be an excitation in $c$ if the excitation was transferred in the previous two steps collectively and thus we are guaranteed to have successfully added a symmetric excitation. 
\end{enumerate}

Only the steps (c) and (e) require non-local interactions mediated by the waveguide. The rest of the steps just require $\pi$-pulses with either individual and collective drivings. For simplicity in the discussion, we assume that they can be done without introducing additional errors. In practical terms, this requires that the modulus and phase of the Rabi coupling can be controlled on the level of $(N P_\oned)^{-1}$ to achieve the desired scaling. Then, the probability of successfully heralding the excitations, $p_{m\rightarrow m+1}$, is the product of the probability in the steps (c) and (e), and the overall error, $I_{m\rightarrow m+1}$, is approximately the sum of the errors in each step.

Finally, to achieve $m$ excitations one needs $m$ successful heraldings and in case of a failed detection, one must start over in $\ket{\phi_0}_\rt$. This implies that the overall probability scales exponentially with $m$ as $p_m=\prod_{k=0}^{m-1} p_{k\rightarrow k+1}$, whereas the error is approximately the sum in each step, $I_m \approx \sum_{k=0}^{m-1} I_{k\rightarrow k+1}$. 

Now, we proceed to analyze the probabilities and errors in steps (c) and (e), obtaining analytical approximations and numerically checking them with exact master equation simulations.

\textbf{Success probabilities in steps (c-e)}

The two steps (c) and (e) are based on Quantum Zeno dynamics \cite{facchi02a} and can be described as particular cases of a more general situation that we consider in Appendix \ref{appendix:zeno}, to which we refer the interested reader for full details of the calculation. To provide an intuitive picture on how the calculation works, we illustrate the main points studying step (c).

In step (c), the initial state of the \emph{source/target} atoms, which are the only ones participating in this step, is $\ket{e_1}_\rs \ket{\phi_m^{\rs_1}}_\rt$. Then, we switch on the laser $\Omega^\trg_{ce_1}$ in the target ensemble and let the system evolve under the evolution of the laser plus waveguide dynamics. The effective non-hermitian Hamiltonian is given in this case by:
\begin{align}\label{eq:effH}
H_\mathrm{eff} 
	=& \frac{1}{2} \Omega_{ce_1}^\trg  \left( S^\trg_{ce_1}+S^\trg_{e_1c} \right)  \\
	 & -\mathrm{i} \frac{\Gamma_\oned}{2}
	\big(\sigma_{e_1g}^\src + S_{e_1g}^\trg \big) \big(\sigma_{ge_1}^\src + S_{ge_1}^\trg  \big)
	-\mathrm{i} \frac{\Gamma^*}{2}\sum_{n}\sigma^{n}_{e_1e_1}. \notag
\end{align}

It can be easily shown that, $H_\mathrm{eff}$, only connects the initial state with $\ket{g}_\rs S_{e_1 g}^\trg \ket{\phi_m^{s_1}}_\rt$ and $\ket{g}_\rs S_{c g}^\trg \ket{\phi_m^{s_1}}_\rt$, such that its evolution can be easily obtained for all parameter regimes. Furthermore, when $\Omega_{ce_1}^\trg \ll N_m\Gamma_\oned$ [with $N_m=N-m$], the dynamics is constrained by the strong dissipation to move within the so-called decoherence-free subspace, which in this case is given by the states, 
$\{
\sqrt{\frac{N_m}{N_m+1}}\ket{e_1}_\rs \ket{\phi_m^{s_1}}_\rt 
	-\sqrt{\frac{1}{N_m+1}}\ket{g}_\rs S_{e_1 g}^\trg \ket{\phi_m^{s_1}}_\rt,
\ket{g}_\rs S_{c g}^\trg \ket{\phi_m^{\rs_1}}_\rt 
\}$.
Using this restricted evolution, one can obtain analytical expressions for the populations of the different states as a function of time (see Appendix), which can be compared to the exact numerical solution. The decay of the probabilities is attributed to two mechanisms: on the one hand, the emission of free-space photons, occurring at a rate $\Gamma^*$, and on the other hand, 
there is always a small population in the superradiant state, which decays at a rate proportional to $\Gamma_{\mathrm{sp}} = \frac{N_m \Omega_{ce_1}^{\trg 2}}{(N_m+1)^2 \Gamma_{\oned}}$. It can be shown that the optimal $\Omega^\trg_{c e_1}$ to minimize errors is the one for which $\Gamma_{sp} =  \Gamma^*$, that is, $\Omega_{ce_1}^\trg \approx \sqrt{N_m\Gamma_\oned \Gamma^*}$. With this choice, the population of the desired state, $\ket{g}_\rs S_{c g}^\trg \ket{\phi_m^{\rs_1}}$, is maximized at time, $T^c_\mathrm{opt}=\frac{\pi}{\sqrt{\Gamma_\oned\Gamma^*}}$. This gives us the probability of succeeding in this step for a given $P_\oned$ and $N_m$, which reads:
\begin{equation} \label{eq:pc}
p_c(T^c_\mathrm{opt})
	=\frac{N_m}{N_m+1}e^{-\frac{\pi}{\sqrt{P_\oned}}}\,.
\end{equation}

\begin{figure}[t]
	\centering
	\includegraphics[width=0.4\textwidth]{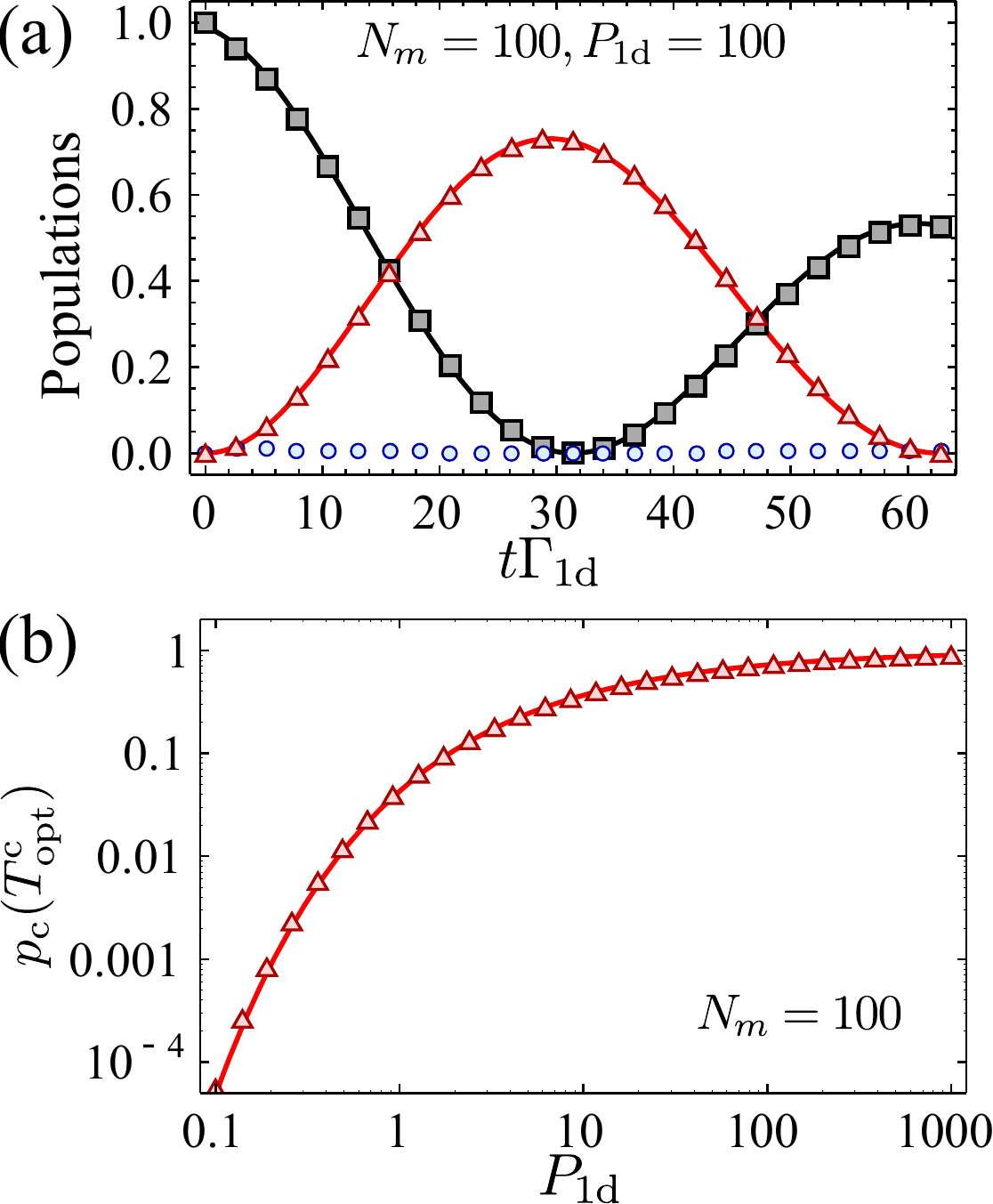}
	\caption{
		(a) Exact population dynamics of step (c) of protocol 1 of the states $\{\ket{e_1}_\rs \ket{\phi_m^{s_1}}_\rt,\ket{g}_\rs S_{e_1 g}^\trg \ket{\phi_m^{s_1}}_\rt, \ket{g}_\rs S_{c g}^\trg \ket{\phi_m^{s_1}}_\rt\}$ in black squares, blue circles and red triangles respectively for $N_m=P_\oned=100$. In black/red solid lines we compare with the approximated results from the restricted evolution obtained under Quantum Zeno dynamics. (b) Optimal probability of step (c) of protocol 1 as a function of $P_\oned$ for $N_m=100$. The markers correspond to the exact numerical evolution, whereas the solid line is the analytical expression of Eq.~(\ref{eq:pc}) obtained under Quantum Zeno Dynamics.
	}
	\label{fig:numerical}
\end{figure}

In Fig.~\ref{fig:numerical} we compare the populations and optimal probabilities obtained by solving the complete (effective) Hamiltonian evolution of Eq.~\ref{eq:effH} and by solving the approximated evolution obtained under Quantum Zeno dynamics. We observe that the agreement between the two approaches in the regime of parameters we are considering is excellent. Moreover, we observe that $p_c\approx 1$ when $P_\oned\gg 1$.

The mechanism behind step (e) is the same as in step (c), just interchanging the role of the \emph{source/target} atoms by the \emph{target/detector} ensemble, which results only in small differences in the prefactors. Thus, to avoid unnecessary repetitions, we write the optimal probability in this case without giving more details here, which reads
\begin{equation} \label{eq:pe}
p_e(T^e_\mathrm{opt})
	= \frac{N_\rd}{N_\rd + m+1} e^{- \pi / \sqrt{(m+1) P_\oned}}\,.
\end{equation}

The interested reader can go to Appendix~\ref{appendix:zeno} where these steps were considered in full generality. Summing up, the overall success probability to go from $m$ to $m+1$ scales as:
\begin{equation} \label{eq:probI}
p_{m\rightarrow m+1}
	= p_c \cdot p_e
	\geq \frac{N_\rd}{N_\rd + m+1}\frac{N_m}{N_m+1}e^{-\frac{2 \pi}{\sqrt{P_\oned}}}\,.
\end{equation}

\textbf{Errors in steps (c-e)}

\begin{figure}[t]
	\centering
	\includegraphics[width=0.4\textwidth]{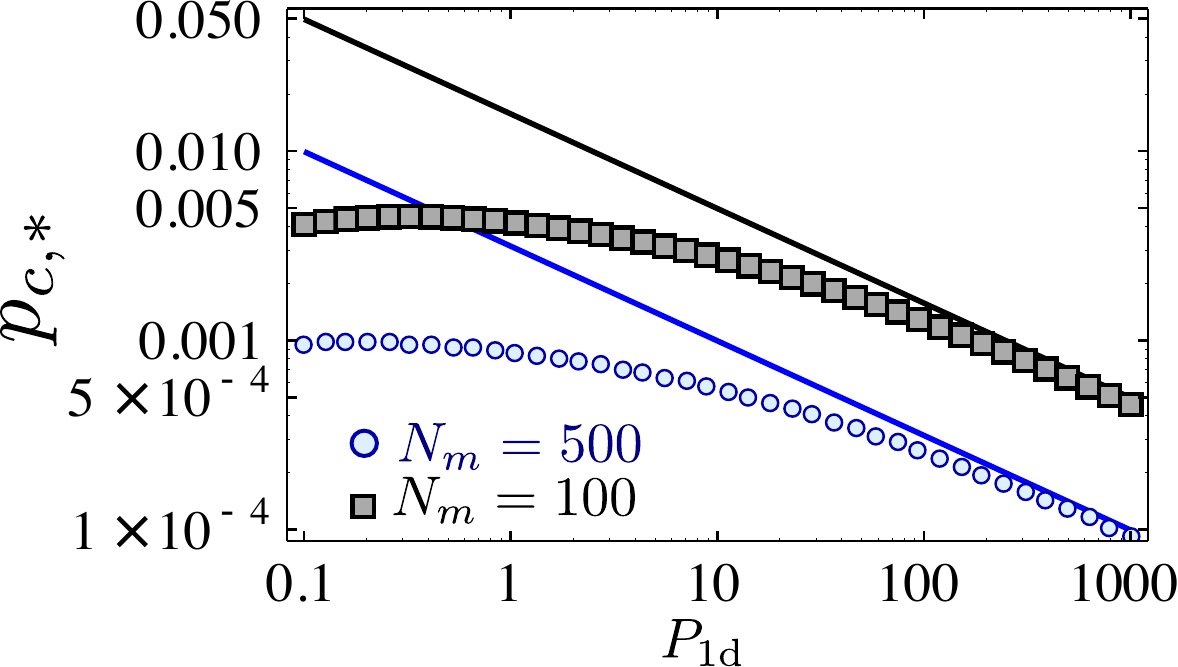}
	\caption{(a): Exact (markers) and estimated upper bound (solid lines) of the errors from spontaneous emission, $p_{c,*}$, as a function of the Purcell Factor $P_\oned$ for different $N_m = N-m$ and fixed $\alpha=1$. The exact calculation is obtained from integrating populations in Eq.~\ref{eq:pjump}, whereas for the analytical upper bound we use Eq.~(\ref{eq:pjumpeff}). }
	\label{fig:pjump}
\end{figure}

As we explained in the overview, in case of unsuccessful detection one needs to pump back all the excitations in the \emph{target} ensemble and restart the protocol from the beginning. Thus, it is easy to check that the only possible errors which cannot be corrected through post-selection in this protocol are individual quantum jumps in the $e_1-c$ transition of the target ensemble, which we assume to be supressed by a factor $\alpha$. These errors would appear in step (c) when the target ensemble is excited. The individual or collective quantum jumps in any one of the other transitions is incompatible with the detection of population in level $c$ of the \emph{detector} ensemble at the final step and, as such, they will not contribute to the infidelity. Using Eq.~(\ref{eq:unravel}), one can calculate the probability of emitting a leaky photon from $e_1 $ to $c$ in step (c) to be
\begin{align} \label{eq:pjump}
p_{c,*}
	\approx \alpha \Gamma^*\int_0^{T_\mathrm{opt}^c} \rd t |c_{e,\mathrm{target}} (t) |^2
\end{align}
where the main contribution comes from spontaneous emission while the laser $\Omega_{ce_1}^\trg$ is switched on. The population in the excited state of target ensemble, $\ket{g}_\rs S_{e_1 g}^\trg \ket{\phi_m^{s_1}}_\rt$, which we denote by $|c_{e,\mathrm{target}} (t) |^2$, can be calculated either with the exact numerical evolution, or using the analytical approximation obtained within the Quantum Zeno formalism. Further contributions to $p_{c,*}$ arise from spurious excitations remaining in $e_1$ after the laser is switched off, but can be neglected as we show in the appendix. Note, that we have already taken into account the reduction factor $\alpha$ of the spontaneous emission in this optical transition. We finally obtain
\begin{align} \label{eq:pjumpeff}
p_{c,*}
	&\approx \alpha \frac{\pi}{2 N_m \sqrt{P_\oned} }\,.
\end{align}

In Fig.~\ref{fig:pjump} we plot the probability $p_{c,*}$ obtained from an exact numerical calculation using Eq.~(\ref{eq:pjump}) and compare it with the asymptotic expression of Eq.~(\ref{eq:pjumpeff}) obtained within the Quantum Zeno dynamics, showing that the agreement is excellent when $P_\oned,N_m\gg 1$. The error introduced when loading a single collective excitation into the target ensemble is then $I_{m \rightarrow m+1} \approx p_{c,*}$.

This can reach the desired scaling when the spontaneous emission of this $c-e_1$ transition can be strongly suppressed by a factor of $\alpha \sim 1/\sqrt{P_\oned}$. Becaus there are several proposals on how to implement this supression \cite{porras08a,borregaard15a}, we only shortly discuss how a two-photon transition (as sketched in Fig.~\ref{fig:Scheme}c) can be used.
In this case, the state $c$ is connected to the excited state $e_1$ through a two-photon transition via the intermediate state $a$. The level $c$ is chosen, e.g., such that the direct transition $c \leftrightarrow e_1$ is dipole forbidden. Then, denoting as $\Omega_1 (\Omega_2)$ the Rabi 
frequency connecting $\ket{c}\leftrightarrow \ket{a} (,\ket{a}\leftrightarrow \ket{e_1})$, and choosing a detuning $\Delta_a\gg \Omega_1,\Omega_2$, the microwave and Raman laser induce a two photon transition between $c \leftrightarrow e_1$ with effective driving $\Omega_{a}=\frac{\Omega_1\Omega_2}{4\Delta_a}$ and a reduced emission rate $\Gamma_\mathrm{c}=\alpha\Gamma*$, where the supression $\alpha = \frac{|\Omega_1|^2}{4 \Delta_a^2} \ll 1$ can be very small.

\textbf{Summary of the figures of merit of protocol 1}

Summarizing, for $N_\rd,N, P_\oned \gg 1$ the success probability and error to load a single symmetric excitation in the \emph{target} ensemble is given by
\begin{align}
p_{m \rightarrow m+1} 
	\approx \ee^{-2 \pi / \sqrt{P_\oned}} \ \mathrm{and} \
I_{m\rightarrow m+1} 
	=p_{c,*}\,,
\end{align}
The total success probability $p_m = \prod_{j=1}^{m-1} p_{j \rightarrow j+1} $ and infidelity 	$I_{m} \approx \sum_{j = 1}^{m-1} I_{m\rightarrow m+1}$ are then given by
\begin{align}
	p_m \approx  \ee^{-2 \pi m / \sqrt{P_\oned}} \ \mathrm{and} \
	I_m \approx m p_{c,*}.
\end{align}

As we explained initially, the goal is to obtain, at least, the same error scaling in the generation of the atomic excitations as in the subsequent atom-photon mapping, that is, $\varepsilon_\mathrm{ph} \sim \frac{m}{N P_\oned}$. Thus, it is enough that the cancellation of spontaneous emission in the $c-e_1$ transition is reduced by a factor $\alpha \sim 1/\sqrt{P_\oned}$.

\subsection{Protocol 2: Excitations added using quantum memories \label{sec:protocol2}}

Even though the overall probability, $p_m$, of protocol 1 can be large for systems with $P_\oned,N\gg 1$, it is still scaling exponentially with the photon number $m$. In protocol 2, we take advantage of the existence of several hyperfine levels, that can work as quantum memories, to design a protocol which circumvents the exponential scaling~\cite{gonzaleztudela17a,fiurasek05a}.

The main difference is that instead of \emph{loading} the collective excitations into the same hyperfine level $s_1$, we store them in different ones $\{s_n\}$, and them combine them a posteriori using post-selection in a tree-like structure (see below). The key to avoid exponential scaling is that in case of unsuccessful heralding, the coherence of the excitations is not heavily affected such that one can try to load it again without starting from the very beginning.

\textbf{Atom-waveguide resources}

The atomic level structure that we require for this protocol is similar to the one used in protocol 1 [see Fig.~\ref{fig:Scheme}], with the extra requirement of a set of hyperfine levels, $\{s_n\}_n$, in which one can store collective excitations. It is important to emphasize here that these levels can also be different vibrational states of the atoms, which enlarges the set of available states to store excitations. Another advantage with respect to the previous protocol is that a single \emph{detector} atom, instead of an ensemble, suffices.
 
\textbf{Overview of the protocol}

Here we present a general overview of the protocol, which has several differences with respect to protocol 1 to be able to reach the desired scaling of infidelities. The first one is that to analyze the \emph{loading} of single collective excitations we assume to start in a state $\ket{\phi_m^{ \{ s_n\}}}_\rt$ of the target ensemble, where $m$ excitations are stored in several metastable states, denoted as $ \{ s_n\}$. Starting with this state, the protocol works as follows:
\begin{enumerate}
	\item [(a)]
	The first step is the same as in the previous protocol, but the transfer of population in the \emph{target} ensemble is not needed as the excitations are already stored in different states. Thus, we only change the state of the detector atom to decouple it from the waveguide dynamics in this step, i.e.,
	\begin{equation}
	\ket{c}_\rs  \ket{\phi_m^{ \{ s_n\}}}_\rt \ket{s}_\rd
	\rightarrow
	\ket{c}_\rs \ket{\phi_m^{ \{ s_n\}}}_\rt \ket{s_1}_\rd.
	\end{equation}
	\item [(b)]
	Next, the source atom is excited with a fast $\pi$-pulse,
	\begin{equation}
	\ket{c}_\rs \ket{\phi_m^{ \{ s_n\}}}_\rt \ket{s_1}_\rd
	\rightarrow 
	\ket{e_1}_\rs \ket{\phi_m^{ \{ s_n\}}}_\rt \ket{s_1}_\rd\,.
	\end{equation}
	
	\item [(c)] Immediately after that, the laser $\Omega_{ce_1}^\trg$ in the \emph{target} ensemble is switched on to trigger the transfer of excitation under Quantum Zeno Dynamics, i.e.
	\begin{equation}
	\ket{e_1}_\rs \ket{\phi_m^{ \{ s_n\}}}_\rt \ket{s_1}_\rd
	\rightarrow
	\ket{g}_\rs S_{cg}^{(\rt)}\ket{\phi_m^{ \{ s_n\}}}_\rt \ket{s_1}_\rd.
	\end{equation}
	This is actually the same step as in protocol 1.
	
	\item [(d)] Now, we change again the state of the detector atom, such that it can interact with the waveguide photons and decouple the source atom as
	\begin{equation}
	\ket{g}_\rs S_{cg}^{(\rt)}\ket{\phi_m^{ \{ s_n\}}}_\rt \ket{s_1}_\rd
	\rightarrow
	\ket{c}_\rs S_{cg}^{(\rt)}\ket{\phi_m^{ \{ s_n\}}}_\rt \ket{s}_\rd.
	\end{equation}
	
	\item [(e)] To trigger the change of the detector atom, one needs to modify the previous protocol as otherwise the errors from spontaneous emission events are too large as we explain below. After exciting the target ensemble from $c$ to $e_2$, we use fast $\pi$-pulses between the $c - e_2$ transitions of both the target and detector ensemble instead of relying on the slow Quantum Zeno Dynamics. This switches the interaction through the collective dissipation effectively on and off. The consequence is a reduction in the success probability, but also in the errors. We obtain
	\begin{equation}
	\ket{c}_\rs S_{c g}^{(\rt)}\ket{\phi_m^{ \{ s_n\}}}_\rt \ket{s}_\rd
	\rightarrow
	\ket{c}_\rs \ket{\phi_{m+1}^{ \{ s'_n\}} }_\rt \ket{c}_\rd,
	\end{equation}
	where the $\{ s'_n\}$ are the same as the $\{ s_n\}$ with one additional collective excitation in level $s$.
	
	\item [(f)] The measurement step is the same as in the previous protocol and we denote the probability for successfully heralding as $q_{m \rightarrow m+1}$.
	
	\item [(g)] This protocol has an extra step in case of unsuccessful heralding: one has to apply a \emph{repumping} scheme before starting again with the first step. This has to be done carefully to avoid introducing extra errors that will spoil the fidelities.
	
	\item [(h)] Finally, to avoid the exponential scaling the excitations stored in $\{s_n\}$ have to be combined in a tree-like structure that circumvents the exponential scaling as we show below.
\end{enumerate}

As in the previous protocol, we assume that all the local operations can be performed perfectly, and only consider the errors and success probabilities of the non-local operations induced by the waveguide.

\textbf{Success probabilities of step (e)}

As mentioned, step (c) is the same as in protocol 1, which we already analyzed in detail there. Thus, we only analyze here the new step (e). In this step, we first apply a $\pi$-pulse on the \emph{target} ensemble with $\Omega_{ce_2}^\trg$, which excites the target ensemble as $S_{c g}^{(\rt)}\ket{\phi_m^{ \{ s_n\}}}_\rt \ket{s}_\rd\rightarrow S_{e_2 g}^{(\rt)}\ket{\phi_m^{ \{ s_n\}}}_\rt \ket{s}_\rd$. As the \emph{source} atom is decoupled, we don't write it for simplicity. Then, we leave the system to evolve under the interaction induced by the waveguide, which is given by the following effective Hamiltonian
\begin{equation}\label{eq:Heff2}
H_{\mathrm{eff}}
	= - \frac{\ii\Gamma_\oned}{2} \left(S_{e_2 s}^\trg+\sigma_{e_2 s}^\dtc \right)\left(S_{s e_2}^\trg +\sigma_{s e_2}^\dtc \right)  - \frac{\ii\Gamma^*}{2} \sum_{n}\sigma_{e_2 e_2}^n\,.
\end{equation}

It can be easily shown, that this Hamiltonian only couples $\ket{\psi_1} \propto S_{e_2 g}^{(\rt)}\ket{\phi_m^{ \{ s_n\}}}_\rt \ket{s}_\rd$ and $\ket{\psi_2} = \ket{\phi_{m+1}^{ \{ s'_n\}} }_\rt \ket{e_2}_\rd$, such that its evolution can be analytically obtained as
\begin{align} \label{eqSM:probabilities2}
|\psi_1(t)|^2 
	&=\frac{1}{4}e^{-\Gamma^* t} \Big[ 1+e^{-\Gamma_\oned t}\Big]^2\,\\
|\psi_2(t)|^2 
	&=\frac{1}{4}e^{-\Gamma^* t} \Big[ 1-e^{-\Gamma_\oned t}\Big]^2\,,
\end{align}

To stop the evolution, we apply a fast $\pi$-pulse on the $c-e_2$ transition of both the target and detector ensemble, which transfers any population in $e_2$ to $c$. As the interaction time we choose a time which leads to the correct error scaling. It turns out that this is achieved for $T_f = \Gamma_\oned^{-1}$ at which point the probability of success of this step is given by
\begin{align}
|\phi_2(T_f)|^2 
	=& \frac{(e-1)^2}{4e^2}\big(1-\frac{1}{P_\oned}\big) 
	\approx 0.1\big(1-\frac{1}{P_\oned}\big)\,.
\end{align}

Thus, the \emph{success probability} in heralding a single excitation in this protocol is given by the combination of step (c) and (e), that is,
\begin{align}
q_{m\rightarrow m+1}
	\approx  0.1\big(1-\frac{1}{P_\oned}\big) \frac{N_m}{N_m+1}e^{-\frac{\pi}{\sqrt{P_\oned}}}\,.
\end{align}

\textbf{Errors}

In case of successful heralding, the only possible error is the one that we already considered in protocol 1, that is, incoherent emission in the $e_1-c$ transition, which leads to an error $p_{c,*}=\frac{\alpha}{N\sqrt{P_\oned}}$. However, in case of unsuccessful heralding, as we want to still use the remaining excitations stored in $\{s_n\}$, one must keep track of all the possible errors that can accumulate in the failed attempts before succeeding.

There are three possible sources of errors that accumulate in each attempt. The first one is the incoherent emission in step (c), which we already showed in the previous protocol scales as $p_{c,*}\propto \frac{1}{N\sqrt{P_\oned}}$. The second one is the probability of spontaneous emission in the target ensemble in step (e), which can be calculated analytically using
\begin{align}
p_{e,*}
	=\Gamma^* \int_{0}^{T_f} \dd t |\phi_1(t)|^2  
	\approx 0.67 P_\oned^{-1}.
\end{align}

Note, that we have chosen $T_f$ in such a way that $p_{e,*}\propto 1/P_\oned$ which is important to obtain the desired scaling of errors, $\epsilon \approx \frac{m}{N P_\oned}$. This is the reason for modifying step (e) with respect to protocol 1, as when using Quantum Zeno Dynamics the probability of spontaneous emission in this step would scale as $1/\sqrt{P_\oned}$.

\begin{figure*}[t]
	\centering
	\includegraphics[width=0.993\textwidth]{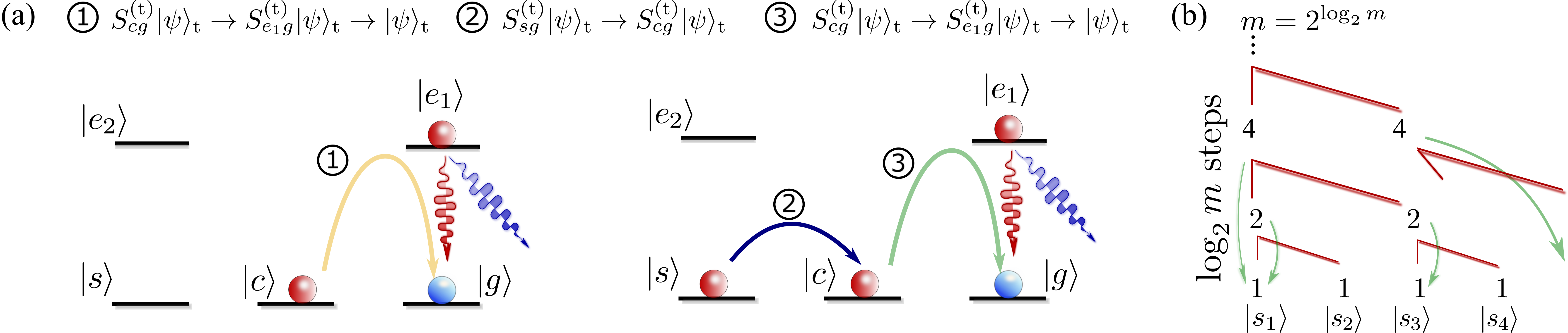}
	\caption{
		(a) The repumping scheme contains three steps for pumping any excitation in the metastable states $c$ or $s$ back to the ground state $g$. (b) Schemes to build up $m$ excitations in a metastable state by beam splitter operations and postselection on zero detection in one of the modes. By doubling the number of excitations at each step one can achieve a subexponential scaling.}
	\label{fig:Repumping}
\end{figure*}

Using the effective non-hermitian evolution of Eq.~\ref{eq:Heff2}, one can also show that the largest quantum jump probability in step (e) is given by the collective quantum jumps in the \emph{target} ensemble, which is $p_{\mathrm{coll}}\approx 0.71$. Thus, if we are able to pump these atoms coherently back to $s$, one can still keep high fidelities in spite of the failed attempt. To do this one must apply a \emph{repumping} procedure in the \emph{target} atoms (depicted in Fig.~\ref{fig:Repumping}) to take advantage of the collective interactions induced by the waveguide:
\begin{enumerate}
	\item
	We apply a $\pi$-pulse on the $c \leftrightarrow e_1$ transition and let the excitation decay superradiantly to $g$.
	\item 
	Once we have made sure that no excitation remains in $c$, we move any excitation in $s$ to $c$.
	\item 
	And finally the first step is repeated and all excitations should have ended up in $g$.
\end{enumerate}

The leading order probability of a spontaneous emission event during this repumping process is
\begin{equation}
p_{*,\mathrm{pump}} \propto \frac{1}{N_m P_\oned}.
\end{equation}

Thus, the overall reduction in the fidelity of the collective excitations of the target ensemble due to spontaneous emission events in each failed attempt is then
\begin{equation}
\epsilon_* 
\approx ( p_{c,*}+p_{e,*}+p_{*,\mathrm{pump}})\frac{m}{N}\,,
\end{equation}
where the $m/N$ factor appears because of the overlap of the initial state with the one obtained after a spontaneous emission event. Typically, $p_{e,*}$ is the largest contribution leading to an infidelity scaling with $\propto m/(N P_\oned)$ as we were aiming for. Thus, the final infidelity taking into account the successful detection and the fact that we need to try an average of $1/q_{m\rightarrow m+1}$ times is
\begin{align}
I_{m\rightarrow m+1}
	=p_{c,*}+\frac{\epsilon_{*}}{q_{m\rightarrow m+1} }
\end{align}

\textbf{Combining excitations through post-selection}

The second part of this protocol is to design ways to combine the $m$ excitations distributed over the metastable states  $\{ s_n\}$ of the \emph{target} ensemble to achieve high-photon numbers in a single hyperfine level while avoiding the exponential scaling. At this stage, it is convenient to use the fact that we are interested in the low excitation limit, $m\ll N$, where the Holstein-Primakoff limit \cite{Holstein1940} is valid. This means that we can approximate collective atomic operators by bosonic operators, i.e., $S_{\alpha g}^\trg \approx \sqrt{N} a^\dagger_\alpha$.  We focus on accumulating $m$ excitations in a single state, i.e., a state $(a^\dagger_s)^m\ket{\mathrm{vac}}$.

The tools we are going to use are collective microwave or laser drivings that can generate either displacement operators of a single mode or lead to beamsplitter-like transformations. The former are obtained by a weak external driving between the state of interest and the ground state $g$. The latter by directly applying, e.g., two photon Raman processes between the states (modes) of interest. Furthermore, we can read out the atomic state very efficiently by pumping to an excited state that emits a collective photon through the waveguide in a cyclic transition. With this, one can verify that a mode (i.e., metastable state) is empty, effectively acting as a projection operator $\mathbb{P}_{0_i}$ on mode $i$. Thus, these tools provide a similar set of tools as the ones used in linear optics protocols with the advantage of having the excitations stored in the metastable states $s_i$.

One possible way to combine excitations is to use a tree-like structure as depicted in Fig.~\ref{fig:Repumping}b. The idea is to \emph{double} the number of excitations in each step such that one can reach $m$ excitations in $\log_2 m$ steps and one does not need to start from the beginning if combining excitations fails, but only in the corresponding branch of the tree. The first building block is to study the process that adds up two states with $k$ excitations, i.e., $\ket{k,k} \rightarrow \ket{2k,0}$, for which the success probability (using a $50-50$ beamsplitter) is upper bounded by
\begin{align}
q_k 
	\leq \frac{(2k)!}{2^{2k}(k!)^2}\,.
\end{align}

Under Stirling's approximation, we can see that $q_k \approx 1/\sqrt{\pi k} \rightarrow 0$ for large excitation numbers.  The mean number of operations to arrive to a state with $m$ excitations can be calculated recursively as $R_m = q^{-1}_{m/2} \left( 1 + 2 R_{m/2} \right)$. One can lower and upper bound this number and show that the scaling is
\begin{align}
\label{eq:R_m}
	R_m \sim \sqrt{m}^{\log_2 m}
\end{align}
where we used that $q_{m}<q_{m/2}$, $q_1=1/2$ and that one needs $\log_2 m$ steps to arrive to $m$ excitations. The combination of a logarithmic number of steps with the polynomial decrease of probability leads to the superpolynomial, but subexponential, scaling of $R_m$. It can be shown~\cite{fiurasek05a} that by combining the doubling steps with single-mode coherent displacement operations one can also prepare arbitrary superpositions of single-mode states.

Finally, we present a way of overcoming the superpolynomial scaling by requiring number resolved detection. The idea is that in each doubling step of the tree, to go from $k$ to $2k$, instead of throwing away all the states in which one does not detect zero excitations in the other mode, we keep the states in which the excitation is less than a given fraction of the original excitations, i.e., $\beta k$. Like this, we are certain we have at least $(2-\beta)k$ excitations in the other mode. The price to pay is that the number of steps in three scales in a less favourable way, as we require $\log_{2-\beta}m$ steps to reach $m$ excitations. However, the probability of detecting $\beta$ excitations, given by $s_\beta = \int_{0}^{\beta m} \rd k \frac{1}{\pi \sqrt{k(m-k)}}$, can be made independent of $m$ (e.g., by choosing $\beta=1/2$, this probability is $s_{1/2} \approx 1/3$). Combining these features, the average number of steps 
\begin{equation}
	R_m
	\sim m^{\log_{2-\beta} (2 /  s_\beta)},
\end{equation}
is minimized for $\beta \approx 0.238$, leading to a scaling as $R_m \sim m^{3.86}$. Thus, by using number resolved detection one can obtain a \emph{polynomial} scaling in the mean number of operations (see also \cite{motes16}), which is a big improvement, especially if one wants to scale the protocol to very large excitation numbers.

\textbf{Summary of the figures of merit of the protocol}

To summarize this protocol, we have shown how for systems with $N,P_\oned\gg 1$, the successful loading of single collective excitations can be done with probability and infidelity scaling (on average) as
\begin{align}
	q_{m\rightarrow m+1} &\approx 0.1 \ee^{- \pi / \sqrt{P_\oned}} \\
		I_{m\rightarrow m+1} &\lesssim \frac{10 m  \ee^{-\pi / \sqrt{P_\oned}}}{N P_{\oned}},
\end{align}

The total success probability to achieve $m$ excitations, $q_m\sim R_{m}^{-1}$ depends in this case on the way one merges the atomic excitations. One is able to obtain either a polynomial or superpolynomial scaling depending on whether number-resolved excitations is used or not, which also determines the total infidelity that will be proportional to $I_m\propto R_m /(NP_\oned)$.

\subsection{Protocol 3: Excitations added directly with two guided modes \label{sec:new}}

We conclude this Section presenting a protocol which avoids the requirement of a transition with strong supression of spontaneous emission in a specific transition, at the expense of using an extra waveguide mode. The key point consists of transferring the excitation from the \emph{source} to the \emph{target} ensemble, while inducing a change in the detector atom in a single step. This single-step transfer can be used either to add excitations directly to the same hyperfine level or to store them in several quantum memories to afterwards join them as described in protocol 1 and 2. To avoid unnecessary repetitions, we will focus on the later case; that is, we start in $\ket{\Phi_m^{\{s_n\}}}_\rt$ and only discuss the heralded loading of a single excitation.

\textbf{Atom-waveguide resources}

\begin{figure}[t]
	\centering
	\includegraphics[width=0.3\textwidth]{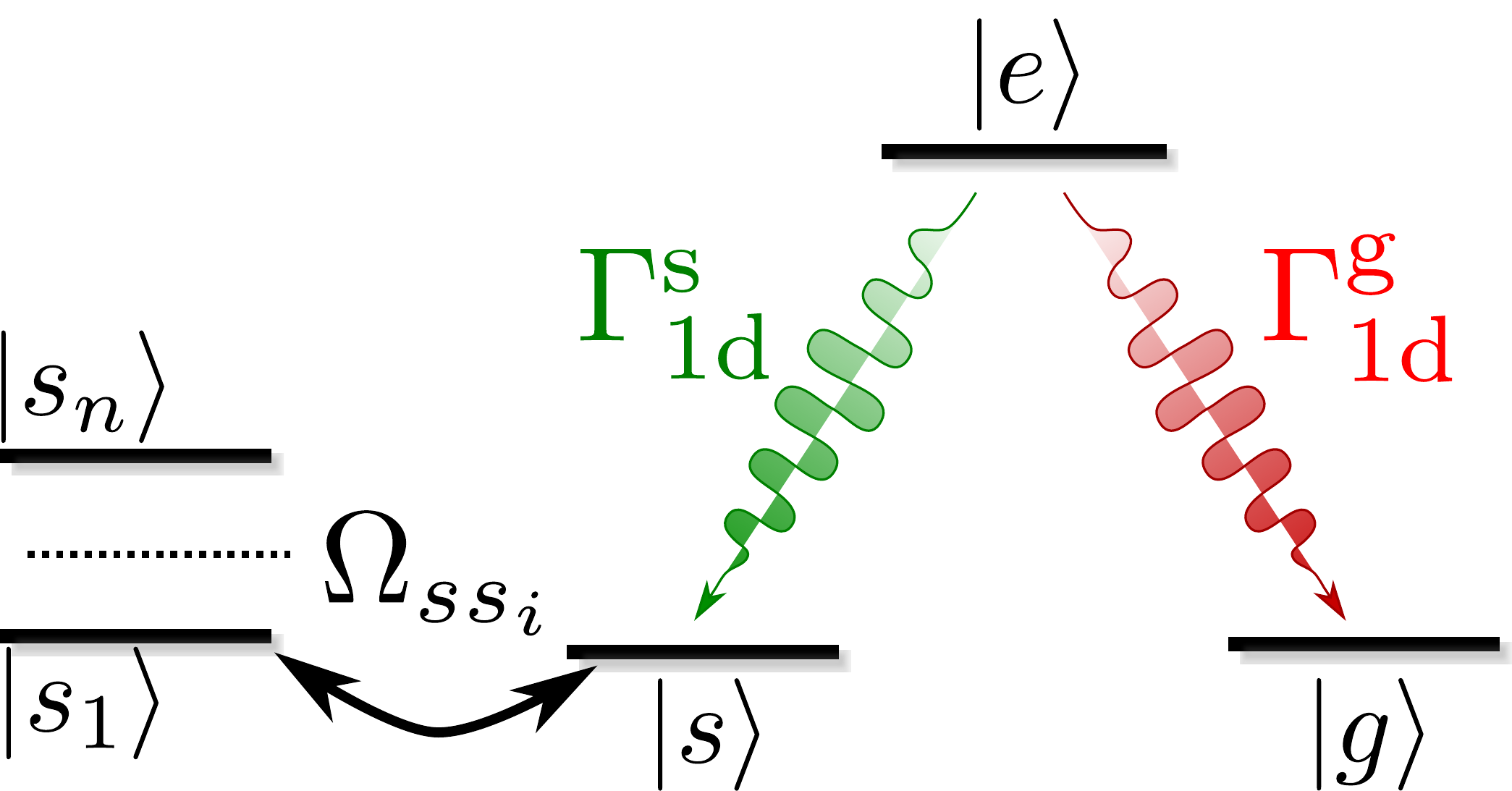}
	\caption{
		(a) Level structure required for protocol 3: $\Lambda$ system for which two different waveguide modes are coupled to the transitions $s-e$, $g-e$, with rates $\Gamma_\oned^{s,g}$ respectively.
	}
	\label{fig:SingleStep}
\end{figure}

The atom-waveguide configuration required for this protocol is depicted in Fig.~\ref{fig:SingleStep}: a $\Lambda$-system, in which two different waveguide modes are coupled to the $s-e$, $g-e$, with rates $\Gamma_\oned^{s/g}$ respectively. We also need auxiliary levels, $\{s_n\}$, to store the excitations. Moreover, we require that the $s$-state of the \emph{source} and the $g$-state of the \emph{detector} can be shifted out of resonance so that unwanted transitions are avoided.

\textbf{Overview of the protocol}

The use of a single-step transfer reduces substantially the complexity of the protocol, which now consists only of the following steps:
\begin{itemize}
	 \item [(a)] The initial state is $\ket{g}_\rs \ket{\phi_m^{ \{ s_n \}}}_\rt \ket{s}_\rd$. First, we excite the source atom by a fast $\pi$-pulse with $\Omega^\src_{ge}$, i.e.,
	 \begin{equation}
	  \ket{g}_\rs \ket{\phi_m^{ \{ s_n \}}}_\rt \ket{s}_\rd
	  \rightarrow 
	  \ket{e}_\rs \ket{\phi_m^{ \{ s_n \}}}_\rt \ket{s}_\rd\,.
	 \end{equation}
	
	 \item [(b)] Just after that, we switch on $\Omega_{ge}^\dtc$ under the Quantum Zeno conditions. By choosing an appropriate time, the excitation is collectively transfered to the \emph{target} ensemble, while inducing a change in the \emph{detector} atom as
	 \begin{equation}
	  \ket{e}_\rs \ket{\phi_m^{ \{ s_n \}}}_\rt \ket{s}_\rd
	  \rightarrow 
	  \ket{g}_\rs S_{sg}^\trg \ket{\phi_m^{ \{ s_n \}}}_\rt \ket{g}_\rd\,.
	 \end{equation}
	 
	 \item [(c)] Then, we perform a measurement on the \emph{detector} atom. If we measure an excitation in $g$, we know we have succeeded. If not, we pump back the \emph{target} atoms and try again until we succeed.
	\end{itemize}
	 
\textbf{Success probabilities}

\begin{figure*}[tb]
	\centering
	\includegraphics[width=0.99\textwidth]{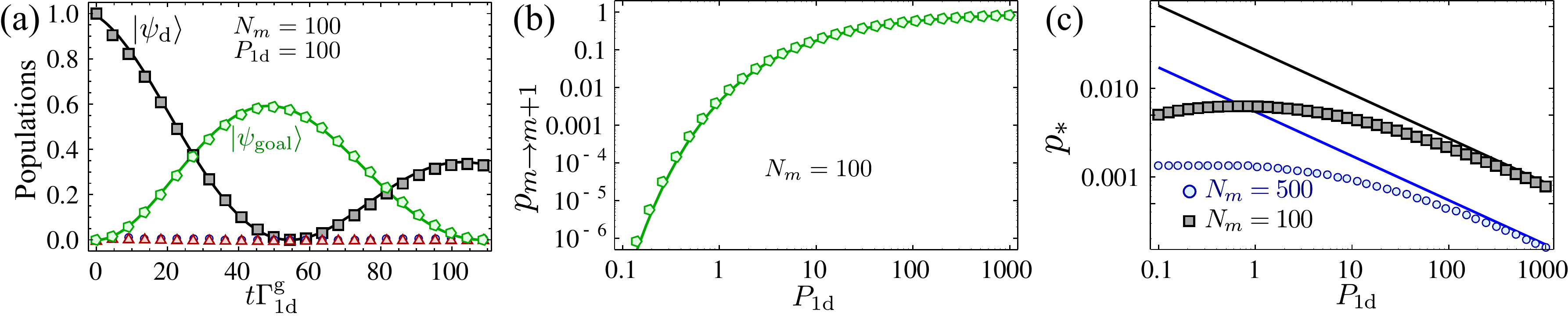}
	\caption{
		(a) Populations for $N_m=100=P_\oned$ calculated with exact non-hermitian Hamiltonian (markers) of Eq.~\ref{eq:3Ham}, together with the analytical approximations (solid lines) within Quantum Zeno dynamics using the optimal $\Omega_{ge}^\dtc = \sqrt{N_m\Gamma_\oned\Gamma^*/3}$. 
		(b) Exact (markers) and approximated (solid line) optimal probability $p_{m \rightarrow m+1}$ obtained at time $T_\mathrm{opt}^\bb \approx \pi\sqrt{3}/\sqrt{\Gamma^*\Gamma_\oned}$. 
		(c) Exact (markers) and assymptotic bound (solid lines) of $p_{*}$ as a function of the Purcell Factor $P_\oned$. The exact calculation is obtained from integrating populations of Eq.~(\ref{eq:3pop}), whereas for the analytical upper bound we use Eq.~(\ref{eq:3sp_em}).
		}
	\label{fig:prot3}
\end{figure*} 

The main part of this protocol is step (b). In this step, the effective non-hermitian Hamiltonian governing the evolution is given by:
\begin{align} \label{eq:3Ham}
H_{\mathrm{eff}} 
	=& \frac{\Omega_{ge}^\dtc}{2}\big(\sigma_{ge}^\dtc+\sigma_{eg}^\dtc \big)
		-\ii \frac{\Gamma_{\oned}^\rg}{2}\big(\sigma_{eg}^\src+S_{eg}^\trg \big)\big(\sigma_{ge}^\src+S_{ge}^\trg \big)\nonumber \\
	& -\ii \frac{\Gamma_{\oned}^\rs}{2}\big(S_{es}^\trg + \sigma_{es}^\dtc \big)\big(S_{se}^\trg + \sigma_{se}^\dtc \big)
		-\ii \frac{\Gamma^*}{2}\sum_{n}\sigma^{n}_{ee}\,.
\end{align}

It can easily be shown (see Appendix for details) that $H_\mathrm{eff}$ only couples the initial state to 3 other states, including the one we aim for, $\ket{\psi_{\mathrm{goal}}} \propto \ket{g}_\rs S_{sg}^\trg \ket{\phi_m^{ \{ s_n \}}}_\rt \ket{g}_\rd$. Therefore, the dynamics can be obtained easily numerically. Moreover, if we work within the Quantum Zeno regime, that is,  $\Omega_d\ll N_m\Gamma_\oned^\rg$, we can show that the dynamics is restricted to the so-called decoherence free-subspace formed by $\ket{\psi_{\mathrm{goal}}}$ and a dark excited state appearing between the three ensembles, that is,
\begin{equation}
\ket{\psi_\rd} 
	\propto \left( \sqrt{N-m} \sigma_{eg}^\src - S_{eg}^\trg + \sigma_{eg}^\dtc \right) 
		\ket{g}_\rs \ket{\phi_m^{ \{ s \}}}_\rt \ket{s}_\rd\,.
\end{equation}

Remarkably, this dark state is independent of the ratio of $\Gamma_\oned^{g,s}$, unlike the two orthogonal superradiant states that appear, which depend on that ratio. Moreover, the overlap of $\ket{\psi_d}$ with the initial state is large, so that large probabilities can be obtained. Using this restricted two dimensional Hilbert space formed by $\ket{\psi_\rd}$ and $\ket{\psi_{\mathrm{goal}}}$, it is possible to obtain analytical expressions of the populations and probabilities. There, one can extract that the optimal ratio between the decay rates is $\Gamma_{\oned}^s / \Gamma_{\oned}^g = \frac{N-m+1}{2}$ and the optimal driving strength is $\Omega_{ge}^\dtc = \sqrt{N_m\Gamma_\oned^g \Gamma^*/3}$, which correspond to the situation where the decay coming from leaky photons, $\Gamma^*$, and the one induced by populating the superradiant states is the same. With those parameters, the population of the decoherence-free states is finally given by
\begin{subequations}\label{eq:3pop}
\begin{align}
|c_\rd(t)|^2 
&\approx \frac{N_m}{N_m+2} e^{-\Gamma^*t} \cos^2(\frac{\sqrt{\Gamma_\oned^g \Gamma^*}t}{2\sqrt{3}})\,, \\
|c_\mathrm{goal}(t)|^2
&\approx \frac{N_m}{N_m+2} e^{-\Gamma^*t} \sin^2(\frac{\sqrt{\Gamma_\oned^g \Gamma^*}t}{2\sqrt{3}})\,.
\end{align}
\end{subequations}

Therefore, by choosing a time $T_\mathrm{opt}^\bb=\frac{\pi\sqrt{3}}{\sqrt{\Gamma_{\oned}^g \Gamma^*}}$, one maximizes the probability of the heralded transfer
\begin{align} \label{eq:3prob}
	p_{m\rightarrow m+1}
		\approx \frac{N_m}{N_m+2} e^{-\sqrt{3}\pi/\sqrt{P_\oned}}\,,
\end{align}
where $P_\oned=\Gamma_\oned^g/\Gamma^*$. In Fig.~\ref{fig:prot3}(a-b) we compare the both the population dynamics and optimal probability using an exact numerical calculation (marker) and the analytical approximations of Eqs.~(\ref{eq:3pop}) and (\ref{eq:3prob}) showing an excellent agreement.

\textbf{Errors}

As in the previous protocols, the problematic quantum jumps leading to errors are the ones that appear from spontaneously emitted photons from excited states of the target ensamble, that is, the population of the state $\ket{g}_\rs S_{eg}^\trg \ket{\phi_m^{ \{ s_n \}}}_\rt \ket{g}_\rd$. The biggest contribution to the decay of this state is given by its overlap with $\ket{\psi_d}$, which can be shown to scale with $1/\sqrt{N_m}$. Thus, using Eqs.~(\ref{eq:3pop}) the final probability of spontaneously emitting a photon reads (see Appendix for details):
\begin{align}\label{eq:3sp_em}
p_{*}
	\approx \frac{\pi\sqrt{3}}{2N_m\sqrt{P_\oned}}\,,\,\mathrm{if}\,\, P_\oned\gg 1\,,
\end{align}
which is confirmed numerically in Fig.~\ref{fig:prot3}(c).
 
If we want to recycle the excitations stored in $\{s_n\}$, we have to apply a repumping procedure, which does not introduce further errors. In this case, it can be done also in a single step by pumping collectively the atoms from $s\rightarrow g$ , which moves $S_{sg}\ket{\phi_m}\rightarrow S_{eg}\ket{\phi_m}\rightarrow\ket{\phi_m}$.  Because this process is done through a collective photon the probability of emitting a free space photon is given by $p_{\mathrm{pump},*} \sim \frac{1}{NP_\oned}$. Thus, the final fidelity of the heralded transfer of a single collective excitation is given by: 
\begin{equation} \label{eq:3infid}
I_{m\rightarrow m+1}
	\approx \frac{1}{p_{m \rightarrow m+1}} (p_{*}+p_{\mathrm{pump},*}) \frac{m}{N}
	\approx \frac{m}{N^2 \sqrt{P_\oned}}\,,
\end{equation}
for systems with $P_\oned,N_m\gg 1$. 

\section{Extension to Two-Mode Photonic States \label{sec:twomode}}

Up to now we have developed different methods to prepare an atomic state in one specific metastable state that can be triggered to emit the desired photonic state in a single mode. An exciting prospect is to extend these protocols to generate entangled states of several atomic excitations that can afterwards be triggered to emit, e.g., into orthogonal waveguide modes with different polarization and/or frequencies. This will enlarge the set of photonic states that can be prepared using different guided modes.

We emphasize that the mapping from symmetric Dicke states to photonic states of the waveguide in the low excitation regime can be straightforwardly generalized to the mapping of two-mode Dicke states, $\ket{\phi_{m,n}^{s_{\uparrow / \downarrow}}} \propto S_{s_\uparrow g}^m S_{s_\downarrow g}^n \ket{g^{\otimes N}}$ to a product of photonic states in the two modes. This originates in the fact that the decay operators commute when applied to a symmetric Dicke state, i.e. $ \left[ S_{s_\uparrow g}, S_{s_\downarrow g}\right] \ket{\phi_{m,n}^{s_{\uparrow / \downarrow}}} =0$ when $m,n \ll N$ (see also Reference \cite{porras08a}).

The goal of this Section is to extend our protocols for adding a single excitation $\ket{\psi} \rightarrow a^\dagger \ket{\psi}$ to adding a single excitation over two modes in a heralded way, i.e. 
\begin{equation}
	\label{eq:goal_2modes}
	\ket{\psi} \rightarrow (c_\uparrow a_\uparrow^\dagger  + c_\downarrow a_\downarrow^\dagger ) \ket{\psi}
\end{equation}
 with $|c_\uparrow|^2+ |c_\downarrow|^2 = 1$. For simplicity of notation, we assume that we work in the low excitation regime, in which the Holstein Primakoff Approximation \cite{holstein40a} can be applied, and we approximate the collective spin operators by bosonic operators, i.e., $S_{s_i g} \approx \sqrt{N} a_i^\dagger$. First, we explain the atom-waveguide resources required. Then, we will show how to adapt the heralded generation to the case of adding a superposition, and finally, we discuss how to apply these tools to generate certain photonic states with potential for metrology beyond the Standard Quantum Limit.
 
 \begin{figure}[t]
 	\centering
 	\includegraphics[width=0.5\textwidth]{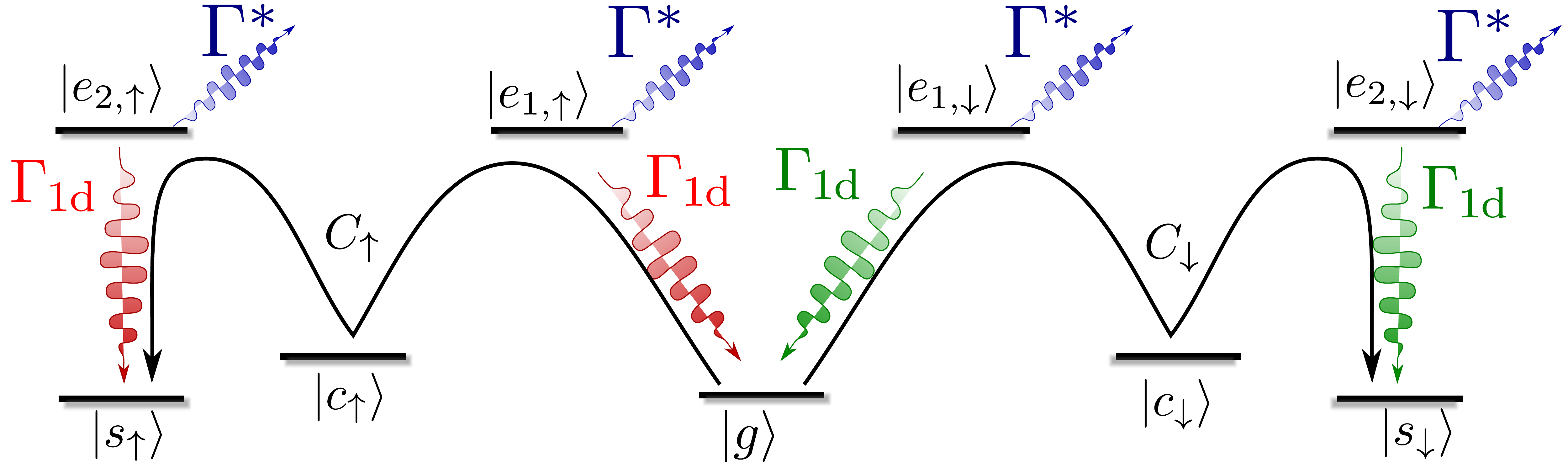}
 	\caption{
 		The level structure for the generation of two-mode states is obtained by mirroring the level structure of Figure \ref{fig:Scheme} around the ground state $g$.
 	}
 	\label{fig:2Modes}
 \end{figure}

\subsection{Atom-waveguide resources}

The generalization of these protocols requires a more elaborate level structure as depicted in Fig.~\ref{fig:2Modes}. It consists of \emph{mirroring} the level structure required for single-mode multiphoton emission required in protocols 1 and 2, in which one side couples to a different guided mode. For simplicity we assume the same decay rate $\Gamma_\oned$ on both sides, though this is not an important point for the proposal.

\subsection{Generalization of Protocols}

For the generalization of the protocols, we focus on directly adding an excitation to the two relevant modes through the waveguide photons. We have shown in Section \ref{sec:single}, that one can perform the following operation: transfer a single excitation from the \emph{source} atom collectively to the $s_{\pm}$ state in the \emph{target} ensemble while inducing a change in the detector atoms, from $s_{\pm}$ to $\ket{c_{\uparrow/\downarrow}}$. We denote this operation as:
\begin{equation}
	\hat{C}_{\uparrow/\downarrow} : 
	\ket{c_{\uparrow/\downarrow}}_\rs \ket{\psi}_\rt \ket{s_{\uparrow/\downarrow}}_\rd 
	\rightarrow	
	\ket{g}_\rs a_{\uparrow/\downarrow}^\dagger \ket{\psi}_\rt \ket{c_{\uparrow/\downarrow}}_\rd 
	+ \ldots \ket{s_{\uparrow/\downarrow}}_\rd,
\end{equation}
where we are not writing explicitly the state of the \emph{source} and \emph{target} ensemble if some error occurs, as we will be able to post-select them with the measurement. For notational simplicity, we write the state of the detector ensemble as if it only contained one emitter, whereas it actually contains $N_\rd$ emitters.  All other possible initial states during the protocol will not change under $C_{\uparrow/\downarrow}$. It is important to emphasize that the measurement is not performed after each $C_{\uparrow / \downarrow}$ but just after both operations have been performed as otherwise the superposition would be destroyed.

The protocol starts by preparing a superposition state in the \emph{source} atom, i.e., $(\alpha_\uparrow \ket{c_\uparrow} + \alpha_\downarrow \ket{c_\downarrow})$. The states in target ensemble are assumed to be in a given state, $\ket{\psi}_\rt$, which may already contain excitations in other hyperfine levels, while the rest are in $g$. The steps of the protocol can be summarized as follows:
\begin{itemize}
	\item [(a)] We start transferring one of the states, e.g., $c_\uparrow$. For that, we prepare the \emph{detector} atoms in $c_\uparrow$ as well, and apply $\hat{C}_\uparrow$ such that
	\begin{align}
		& \left( \alpha_\uparrow \ket{c_\uparrow}_\rs + \alpha_\downarrow \ket{c_\downarrow}_\rs\right) \ket{\psi}_\rt \ket{s_\uparrow}_\rd \nonumber\\
		\stackrel{\hat{C}_\uparrow}{\rightarrow}
		& \alpha_\uparrow \ket{g}_\rs a_\uparrow^\dagger \ket{\psi}_\rt \ket{c_\uparrow}_\rd 
		+ \alpha_\downarrow \ket{c_\downarrow}_\rs \ket{\psi}_\rt \ket{s_\uparrow}_\rd + \ldots \ket{s_\uparrow}_\rd  \nonumber
	\end{align} 
	
	\item [(b)] For the orthogonal mode, we need to transfer the excitation in $s_\uparrow$ to $s_\downarrow$ in the \emph{detector} ensemble to be able to herald the excitation in $c_\downarrow$ at the end. We note, that the popoulation in $c_\uparrow$ of the detector ensemble will not be affected by later transitions. The new state is
	\begin{align}
		&\stackrel{\hat{O}_1}{\rightarrow}
		\alpha_\uparrow \ket{g}_\rs a_\uparrow^\dagger \ket{\psi}_\rt \ket{c_\uparrow}_\rd 
		+ \alpha_\downarrow \ket{c_\downarrow}_\rs \ket{\psi}_\rt \ket{s_\downarrow}_\rd 
		+ \ldots \ket{s_\downarrow}_\rd\,, \notag 
	\end{align} 
	where we used $\hat{O}_1$ to denote this operation. 
	
	\item [(c)] Now, one can apply $\hat{C}_\downarrow$ which results in
	\begin{align}
		&\stackrel{\hat{C}_\downarrow}{\rightarrow}
		 \alpha_\uparrow \ket{g}_\rs a_\uparrow^\dagger \ket{\psi}_\rt \ket{c_\uparrow}_\rd 
		+ \alpha_\downarrow \ket{g}_\rs a_\downarrow^\dagger \ket{\psi}_\rt \ket{c_\downarrow}_\rd 
		+ \ldots \ket{s_\downarrow}_\rd \notag 
 	\end{align} 
 
	\item [ (d)] Now the excitations have been added but each of them are associated to two different metastable states in the detector atoms. In order to be able to post-select them at the same time,  one needs to apply a 50:50 beam splitter between the $c_\uparrow$ and $c_\downarrow$ states of the detector such that
	\begin{align}
 		&\stackrel{\hat{O}_2}{\rightarrow}
		\ket{g}_\rs \frac{1}{\sqrt{2}} \left(\alpha_\uparrow a_\uparrow^\dagger + \alpha_\downarrow a_\downarrow^\dagger\right) \ket{\psi}_\rt \ket{c_\uparrow}_\rd
		+ \ldots \ket{c_\downarrow/s_\downarrow}_\rd \notag 
	 \end{align} 
	 which we denoted as $\hat{O}_2$. After this operation, one can herald the transfer of excitation by measuring the state $g$ of the \emph{detector} atoms as we desired.
	 
	\end{itemize}

We emphasize here again that the measurement has to be performed after all the operations to avoid destroying the superpositions between the ensembles.
Furthermore, we would like to point out that in case no excitation has been detected in $c_\uparrow$, one can still measure the state $c_\downarrow$ of the \emph{detector} atoms, and if this heralding is successful one would generate the state $\frac{1}{\sqrt{2}} \left(\alpha_\uparrow a_\uparrow^\dagger - \alpha_\downarrow a_\downarrow^\dagger\right) \ket{\psi}_\rt$. Depending on the goal state this state could still be useful and one would avoid a reduction in the probability by a factor of 2.
The success probability scales in the same way as for the single mode preparation, as the operation is only applied twice compared to once for single mode states with a factor of 2 for some states.

\subsection{Examples with metrological interest}

One of the main motivations to obtain two-mode multiphoton states of light is the possibility of measuring phases, $\phi$, beyond the limits of classical light \cite{demkowicz-dobrzanski15}. It is well known that classical sources can only achieve the so-called Standard Quantum Limit , i.e., $\Delta \phi \propto \frac{1}{\sqrt{n}}$ with $n$ being the number of photons. However, certain two-mode states of light can show a higher precision, and even reach the Heisenberg scaling, that is, $\Delta \phi \propto \frac{1}{n}$. In this Section, we see how one can obtain some of these states of metrological interests using our protocols.

The simplest states to obtain are the so-called Holland-Burnett states~ \cite{holland93,campos03,cooper10}, which are obtained by applying a beam splitter transformation on a dual Fock state $\ket{\psi_\mathrm{HB}} \propto B \ket{m,m}$, i.e. $n=2m$, which can be shown to achieve a precision given by $\Delta \phi = \frac{1}{\sqrt{n(1+n/2)}}$. As dual Fock states are separable states we can obtain them using our protocols 1-3 to achieve single-mode multiphoton states in the two metastable states $s_{\uparrow/\downarrow}$ separately.

Another class of non-classical states with improved precision are Yurke states, i.e., $\ket{\psi_\mathrm{Yurke}}  \propto \ket{m,m-1} + \ket{m-1,m}$, where $n = 2m$. These states  do not reach the Heisenberg limit, but scale in the same way, i.e., $\Delta \phi = \frac{2}{n}$ \cite{yurke86} (at least when $\phi \approx 0$). We note that the Yurke state can be written as
\begin{equation}
	\ket{\psi_\mathrm{Yurke}}
	\propto  (a_\uparrow^\dagger + a_\downarrow^\dagger) a_\uparrow^{\dagger n-1} a_\downarrow^{\dagger n-1} \ket{0,0}.
\end{equation}
We note here, that the factor of 2 we mentioned for the general protocol can be avoided here because the state $\ket{m,m-1}-\ket{m-1,m}$ can be transformed to the Yurke state by a phase shift operator $\exp(-\ii \pi a_\uparrow^\dagger a_\uparrow)$ (e.g., through driving an optical transition off-resonantly).

As the dual Fock states can be generated efficiently with our protocols (see discussion on Holland-Burnett-States), one only needs to add one single excitation over the two metastable states at the end. Other proposals for the generation of Yurke states using linear optical setups, e.g., by photon substraction \cite{ono16} are also possible by using the transitions between metastable states.

Finally, NOON states i.e.~$\ket{\psi_\mathrm{NOON}} = \ket{n::0} \propto \ket{n,0} + \ket{0,n}$, are the only ones that reach the Heisenberg limit~i.e. $\Delta \phi = \frac{1}{n}$. Using the fundamental theorem of algebra they can be written as follows:
\begin{equation}
	\ket{n::0} 
	\propto (a_\uparrow^{\dagger n} + a_\downarrow^{\dagger n} ) \ket{0,0} 
	= \prod_{j=1}^n (a_\uparrow^\dagger + \ee^{\ii \phi_j} a_\downarrow^\dagger ) \ket{0,0}.
\end{equation}
Thus by adding excitations one-by-one over two modes with the appropriate phase, one can generate these states using the extension of our protocol in an exponential number of steps.
In \cite{kok02}, it is shown, that one can ``double'' a NOON-state, that is joining two states of the form $\ket{m::0}$ to obtain $\ket{2m-2::0}$, heralding on a twofold detector coincidence measurement. By using this method together with the metastable states acting as quantum memories, one can obtain NOON-states in a superpolynomial, but subexponential, mean number of steps. This is very similar to the single-mode scheme: The probability of joining the two states is $\frac{2}{16 \cdot 4^{n-1}} \binom{2n-2}{n-1} \approx \frac{1}{8 \sqrt{\pi(n-1)}}$.
As the phases when adding or doubling states to obtain a NOON-states are very important, the single-mode scheme for a polynomial scaling of the mean number of steps cannot easily be extended to two modes.

\section{Conclusions \& Outlook}\label{sec:outlook}

Summing up, in this work we have revisited the methods discussed in Ref.~\cite{gonzaleztudela17a} to obtain single mode multiphoton states with large photon numbers using atom-waveguide resources. In particular, we have expanded the discussion of the protocols by providing full details of all the subtleties: error sources, repumping procedure, etc. Importantly, we have also provided numerical evidence that the formulas given in ~\cite{gonzaleztudela17a} actually capture the right scaling using exact simulation of the dynamics. Moreover, we have presented a new protocol which simplifies some of the requirements of the previous protocol, i.e., a strong supression of spontaneous emission into a specific state, at the expense of requiring another strongly coupled guided mode. Finally, we have shown how to extend our protocols to directly generate superpositions of atomic excitations that can afterwards be triggered to two-mode entangled states in the waveguide. Though we have focused on the protocols harnessing long-range collective dissipation, one can also export these ideas to other waveguide setups in which one exploits the long-range coherent interactions~\cite{paulisch17a}. An exciting prospect of this work is to combine these protocols with the existence of \emph{free-space subradiant states}~\cite{porras08a,asenjogarcia17a} in which spontaneous emission is highly suppressed to boost even more the fidelities and probabilities of the protocols.

\begin{acknowledgements}
	\textbf{Acknowledgements}
	The work of AGT, VP and JIC was funded by the European Union integrated project \emph{Simulators and Interfaces with Quantum Systems} (SIQS).
	AGT also acknowledges support from Intra-European Marie-Curie Fellowship NanoQuIS (625955).
	VP acknowledges the Cluster of Excellence NIM.
	This work was supported by the ERC grant QUENOCOBA 742102. 
	Funding for HJK was provided by the Office of Naval Research (ONR) Award No. N00014-16-1-2399, by the Air Force Office of Scientific Research (AFOSR) MURI “Photonic Quantum Matter”, by the ONR MURI “Quantum Opto-Mechanics with Atoms and Nanostructured Diamond (QOMAND)”, by NSF Grant No. PHY-1205729, and by the Institute for Quantum Inforation and Matter (IQIM), an NSF Physics Frontiers Center. HJK acknowledges support as a Max Planck Distinguished Scholar that enabled his participation in this collaboration.
\end{acknowledgements}

\bibliography{Refs}

\begin{thebibliography}{47}%
\makeatletter
\providecommand \@ifxundefined [1]{%
 \@ifx{#1\undefined}
}%
\providecommand \@ifnum [1]{%
 \ifnum #1\expandafter \@firstoftwo
 \else \expandafter \@secondoftwo
 \fi
}%
\providecommand \@ifx [1]{%
 \ifx #1\expandafter \@firstoftwo
 \else \expandafter \@secondoftwo
 \fi
}%
\providecommand \natexlab [1]{#1}%
\providecommand \enquote  [1]{``#1''}%
\providecommand \bibnamefont  [1]{#1}%
\providecommand \bibfnamefont [1]{#1}%
\providecommand \citenamefont [1]{#1}%
\providecommand \href@noop [0]{\@secondoftwo}%
\providecommand \href [0]{\begingroup \@sanitize@url \@href}%
\providecommand \@href[1]{\@@startlink{#1}\@@href}%
\providecommand \@@href[1]{\endgroup#1\@@endlink}%
\providecommand \@sanitize@url [0]{\catcode `\\12\catcode `\$12\catcode
  `\&12\catcode `\#12\catcode `\^12\catcode `\_12\catcode `\%12\relax}%
\providecommand \@@startlink[1]{}%
\providecommand \@@endlink[0]{}%
\providecommand \url  [0]{\begingroup\@sanitize@url \@url }%
\providecommand \@url [1]{\endgroup\@href {#1}{\urlprefix }}%
\providecommand \urlprefix  [0]{URL }%
\providecommand \Eprint [0]{\href }%
\providecommand \doibase [0]{http://dx.doi.org/}%
\providecommand \selectlanguage [0]{\@gobble}%
\providecommand \bibinfo  [0]{\@secondoftwo}%
\providecommand \bibfield  [0]{\@secondoftwo}%
\providecommand \translation [1]{[#1]}%
\providecommand \BibitemOpen [0]{}%
\providecommand \bibitemStop [0]{}%
\providecommand \bibitemNoStop [0]{.\EOS\space}%
\providecommand \EOS [0]{\spacefactor3000\relax}%
\providecommand \BibitemShut  [1]{\csname bibitem#1\endcsname}%
\let\auto@bib@innerbib\@empty
\bibitem [{\citenamefont {Gonz\'alez-Tudela}\ \emph {et~al.}(2017)\citenamefont
  {Gonz\'alez-Tudela}, \citenamefont {Paulisch}, \citenamefont {Kimble},\ and\
  \citenamefont {Cirac}}]{gonzaleztudela17a}%
  \BibitemOpen
  \bibfield  {author} {\bibinfo {author} {\bibfnamefont {A.}~\bibnamefont
  {Gonz\'alez-Tudela}}, \bibinfo {author} {\bibfnamefont {V.}~\bibnamefont
  {Paulisch}}, \bibinfo {author} {\bibfnamefont {H.~J.}\ \bibnamefont
  {Kimble}}, \ and\ \bibinfo {author} {\bibfnamefont {J.~I.}\ \bibnamefont
  {Cirac}},\ }\href {\doibase 10.1103/PhysRevLett.118.213601} {\bibfield
  {journal} {\bibinfo  {journal} {Phys. Rev. Lett.}\ }\textbf {\bibinfo
  {volume} {118}},\ \bibinfo {pages} {213601} (\bibinfo {year}
  {2017})}\BibitemShut {NoStop}%
\bibitem [{\citenamefont {Kimble}(2008)}]{kimble08a}%
  \BibitemOpen
  \bibfield  {author} {\bibinfo {author} {\bibfnamefont {H.}~\bibnamefont
  {Kimble}},\ }\href@noop {} {\bibfield  {journal} {\bibinfo  {journal}
  {Nature}\ }\textbf {\bibinfo {volume} {453}},\ \bibinfo {pages} {1023}
  (\bibinfo {year} {2008})}\BibitemShut {NoStop}%
\bibitem [{\citenamefont {Caves}(1981)}]{caves81}%
  \BibitemOpen
  \bibfield  {author} {\bibinfo {author} {\bibfnamefont {C.~M.}\ \bibnamefont
  {Caves}},\ }\href {\doibase 10.1103/physrevd.23.1693} {\bibfield  {journal}
  {\bibinfo  {journal} {Physical Review D}\ }\textbf {\bibinfo {volume} {23}},\
  \bibinfo {pages} {1693} (\bibinfo {year} {1981})}\BibitemShut {NoStop}%
\bibitem [{\citenamefont {Holland}\ and\ \citenamefont
  {Burnett}(1993)}]{holland93}%
  \BibitemOpen
  \bibfield  {author} {\bibinfo {author} {\bibfnamefont {M.~J.}\ \bibnamefont
  {Holland}}\ and\ \bibinfo {author} {\bibfnamefont {K.}~\bibnamefont
  {Burnett}},\ }\href {\doibase 10.1103/PhysRevLett.71.1355} {\bibfield
  {journal} {\bibinfo  {journal} {Phys. Rev. Lett.}\ }\textbf {\bibinfo
  {volume} {71}},\ \bibinfo {pages} {1355} (\bibinfo {year}
  {1993})}\BibitemShut {NoStop}%
\bibitem [{\citenamefont {Giovannetti}\ \emph {et~al.}(2006)\citenamefont
  {Giovannetti}, \citenamefont {Lloyd},\ and\ \citenamefont
  {Maccone}}]{giovannetti04a}%
  \BibitemOpen
  \bibfield  {author} {\bibinfo {author} {\bibfnamefont {V.}~\bibnamefont
  {Giovannetti}}, \bibinfo {author} {\bibfnamefont {S.}~\bibnamefont {Lloyd}},
  \ and\ \bibinfo {author} {\bibfnamefont {L.}~\bibnamefont {Maccone}},\ }\href
  {\doibase 10.1103/PhysRevLett.96.010401} {\bibfield  {journal} {\bibinfo
  {journal} {Phys. Rev. Lett.}\ }\textbf {\bibinfo {volume} {96}},\ \bibinfo
  {pages} {010401} (\bibinfo {year} {2006})}\BibitemShut {NoStop}%
\bibitem [{\citenamefont {Lounis}\ and\ \citenamefont
  {Orrit}(2005)}]{lounis05a}%
  \BibitemOpen
  \bibfield  {author} {\bibinfo {author} {\bibfnamefont {B.}~\bibnamefont
  {Lounis}}\ and\ \bibinfo {author} {\bibfnamefont {M.}~\bibnamefont {Orrit}},\
  }\href@noop {} {\bibfield  {journal} {\bibinfo  {journal} {Rep. Prog. Phys.}\
  }\textbf {\bibinfo {volume} {68}},\ \bibinfo {pages} {1129} (\bibinfo {year}
  {2005})}\BibitemShut {NoStop}%
\bibitem [{\citenamefont {Xu}\ \emph {et~al.}(1999)\citenamefont {Xu},
  \citenamefont {Vu{\v{c}}kovi{\'c}}, \citenamefont {Lee}, \citenamefont
  {Painter}, \citenamefont {Scherer},\ and\ \citenamefont {Yariv}}]{xu99a}%
  \BibitemOpen
  \bibfield  {author} {\bibinfo {author} {\bibfnamefont {Y.}~\bibnamefont
  {Xu}}, \bibinfo {author} {\bibfnamefont {J.}~\bibnamefont
  {Vu{\v{c}}kovi{\'c}}}, \bibinfo {author} {\bibfnamefont {R.}~\bibnamefont
  {Lee}}, \bibinfo {author} {\bibfnamefont {O.}~\bibnamefont {Painter}},
  \bibinfo {author} {\bibfnamefont {A.}~\bibnamefont {Scherer}}, \ and\
  \bibinfo {author} {\bibfnamefont {A.}~\bibnamefont {Yariv}},\ }\href@noop {}
  {\bibfield  {journal} {\bibinfo  {journal} {JOSA B}\ }\textbf {\bibinfo
  {volume} {16}},\ \bibinfo {pages} {465} (\bibinfo {year} {1999})}\BibitemShut
  {NoStop}%
\bibitem [{\citenamefont {Painter}\ \emph {et~al.}(1999)\citenamefont
  {Painter}, \citenamefont {Lee}, \citenamefont {Scherer}, \citenamefont
  {Yariv}, \citenamefont {O'brien}, \citenamefont {Dapkus},\ and\ \citenamefont
  {Kim}}]{painter99a}%
  \BibitemOpen
  \bibfield  {author} {\bibinfo {author} {\bibfnamefont {O.}~\bibnamefont
  {Painter}}, \bibinfo {author} {\bibfnamefont {R.}~\bibnamefont {Lee}},
  \bibinfo {author} {\bibfnamefont {A.}~\bibnamefont {Scherer}}, \bibinfo
  {author} {\bibfnamefont {A.}~\bibnamefont {Yariv}}, \bibinfo {author}
  {\bibfnamefont {J.}~\bibnamefont {O'brien}}, \bibinfo {author} {\bibfnamefont
  {P.}~\bibnamefont {Dapkus}}, \ and\ \bibinfo {author} {\bibfnamefont
  {I.}~\bibnamefont {Kim}},\ }\href@noop {} {\bibfield  {journal} {\bibinfo
  {journal} {Science}\ }\textbf {\bibinfo {volume} {284}},\ \bibinfo {pages}
  {1819} (\bibinfo {year} {1999})}\BibitemShut {NoStop}%
\bibitem [{\citenamefont {Hughes}(2004)}]{hughes04a}%
  \BibitemOpen
  \bibfield  {author} {\bibinfo {author} {\bibfnamefont {S.}~\bibnamefont
  {Hughes}},\ }\href@noop {} {\bibfield  {journal} {\bibinfo  {journal} {Opt.
  Lett.}\ }\textbf {\bibinfo {volume} {29}},\ \bibinfo {pages} {2659} (\bibinfo
  {year} {2004})}\BibitemShut {NoStop}%
\bibitem [{\citenamefont {Rao}\ and\ \citenamefont {Hughes}(2007)}]{rao07a}%
  \BibitemOpen
  \bibfield  {author} {\bibinfo {author} {\bibfnamefont {V.~M.}\ \bibnamefont
  {Rao}}\ and\ \bibinfo {author} {\bibfnamefont {S.}~\bibnamefont {Hughes}},\
  }\href@noop {} {\bibfield  {journal} {\bibinfo  {journal} {Phys. Rev. B}\
  }\textbf {\bibinfo {volume} {75}},\ \bibinfo {pages} {205437} (\bibinfo
  {year} {2007})}\BibitemShut {NoStop}%
\bibitem [{\citenamefont {Laucht}\ \emph
  {et~al.}(2012{\natexlab{a}})\citenamefont {Laucht}, \citenamefont {P{\"u}tz},
  \citenamefont {G{\"u}nthner}, \citenamefont {Hauke}, \citenamefont {Saive},
  \citenamefont {Fr{\'e}d{\'e}rick}, \citenamefont {Bichler}, \citenamefont
  {Amann}, \citenamefont {Holleitner}, \citenamefont {Kaniber},\ and\
  \citenamefont {Finley}}]{laucht12a}%
  \BibitemOpen
  \bibfield  {author} {\bibinfo {author} {\bibfnamefont {A.}~\bibnamefont
  {Laucht}}, \bibinfo {author} {\bibfnamefont {S.}~\bibnamefont {P{\"u}tz}},
  \bibinfo {author} {\bibfnamefont {T.}~\bibnamefont {G{\"u}nthner}}, \bibinfo
  {author} {\bibfnamefont {N.}~\bibnamefont {Hauke}}, \bibinfo {author}
  {\bibfnamefont {R.}~\bibnamefont {Saive}}, \bibinfo {author} {\bibfnamefont
  {S.}~\bibnamefont {Fr{\'e}d{\'e}rick}}, \bibinfo {author} {\bibfnamefont
  {M.}~\bibnamefont {Bichler}}, \bibinfo {author} {\bibfnamefont {M.-C.}\
  \bibnamefont {Amann}}, \bibinfo {author} {\bibfnamefont {A.~W.}\ \bibnamefont
  {Holleitner}}, \bibinfo {author} {\bibfnamefont {M.}~\bibnamefont {Kaniber}},
  \ and\ \bibinfo {author} {\bibfnamefont {J.~J.}\ \bibnamefont {Finley}},\
  }\href {\doibase 10.1103/PhysRevX.2.011014} {\bibfield  {journal} {\bibinfo
  {journal} {Phys. Rev. X}\ }\textbf {\bibinfo {volume} {2}},\ \bibinfo {pages}
  {011014} (\bibinfo {year} {2012}{\natexlab{a}})}\BibitemShut {NoStop}%
\bibitem [{\citenamefont {Arcari}\ \emph {et~al.}(2014)\citenamefont {Arcari},
  \citenamefont {S\"ollner}, \citenamefont {Javadi}, \citenamefont
  {Lindskov~Hansen}, \citenamefont {Mahmoodian}, \citenamefont {Liu},
  \citenamefont {Thyrrestrup}, \citenamefont {Lee}, \citenamefont {Song},
  \citenamefont {Stobbe},\ and\ \citenamefont {Lodahl}}]{arcari14a}%
  \BibitemOpen
  \bibfield  {author} {\bibinfo {author} {\bibfnamefont {M.}~\bibnamefont
  {Arcari}}, \bibinfo {author} {\bibfnamefont {I.}~\bibnamefont {S\"ollner}},
  \bibinfo {author} {\bibfnamefont {A.}~\bibnamefont {Javadi}}, \bibinfo
  {author} {\bibfnamefont {S.}~\bibnamefont {Lindskov~Hansen}}, \bibinfo
  {author} {\bibfnamefont {S.}~\bibnamefont {Mahmoodian}}, \bibinfo {author}
  {\bibfnamefont {J.}~\bibnamefont {Liu}}, \bibinfo {author} {\bibfnamefont
  {H.}~\bibnamefont {Thyrrestrup}}, \bibinfo {author} {\bibfnamefont {E.~H.}\
  \bibnamefont {Lee}}, \bibinfo {author} {\bibfnamefont {J.~D.}\ \bibnamefont
  {Song}}, \bibinfo {author} {\bibfnamefont {S.}~\bibnamefont {Stobbe}}, \ and\
  \bibinfo {author} {\bibfnamefont {P.}~\bibnamefont {Lodahl}},\ }\href
  {\doibase 10.1103/PhysRevLett.113.093603} {\bibfield  {journal} {\bibinfo
  {journal} {Phys. Rev. Lett.}\ }\textbf {\bibinfo {volume} {113}},\ \bibinfo
  {pages} {093603} (\bibinfo {year} {2014})}\BibitemShut {NoStop}%
\bibitem [{\citenamefont {Somaschi}\ \emph {et~al.}(2016)\citenamefont
  {Somaschi}, \citenamefont {Giesz}, \citenamefont {De~Santis}, \citenamefont
  {Loredo}, \citenamefont {Almeida}, \citenamefont {Hornecker}, \citenamefont
  {Portalupi}, \citenamefont {Grange}, \citenamefont {Ant{\'o}n}, \citenamefont
  {Demory} \emph {et~al.}}]{somaschi16a}%
  \BibitemOpen
  \bibfield  {author} {\bibinfo {author} {\bibfnamefont {N.}~\bibnamefont
  {Somaschi}}, \bibinfo {author} {\bibfnamefont {V.}~\bibnamefont {Giesz}},
  \bibinfo {author} {\bibfnamefont {L.}~\bibnamefont {De~Santis}}, \bibinfo
  {author} {\bibfnamefont {J.}~\bibnamefont {Loredo}}, \bibinfo {author}
  {\bibfnamefont {M.~P.}\ \bibnamefont {Almeida}}, \bibinfo {author}
  {\bibfnamefont {G.}~\bibnamefont {Hornecker}}, \bibinfo {author}
  {\bibfnamefont {S.~L.}\ \bibnamefont {Portalupi}}, \bibinfo {author}
  {\bibfnamefont {T.}~\bibnamefont {Grange}}, \bibinfo {author} {\bibfnamefont
  {C.}~\bibnamefont {Ant{\'o}n}}, \bibinfo {author} {\bibfnamefont
  {J.}~\bibnamefont {Demory}},  \emph {et~al.},\ }\href@noop {} {\bibfield
  {journal} {\bibinfo  {journal} {Nat. Photonics}\ } (\bibinfo {year}
  {2016})}\BibitemShut {NoStop}%
\bibitem [{\citenamefont {Reiserer}\ and\ \citenamefont
  {Rempe}(2015)}]{reiserer15}%
  \BibitemOpen
  \bibfield  {author} {\bibinfo {author} {\bibfnamefont {A.}~\bibnamefont
  {Reiserer}}\ and\ \bibinfo {author} {\bibfnamefont {G.}~\bibnamefont
  {Rempe}},\ }\href {\doibase 10.1103/revmodphys.87.1379} {\bibfield  {journal}
  {\bibinfo  {journal} {Rev. Mod. Phys.}\ }\textbf {\bibinfo {volume} {87}},\
  \bibinfo {pages} {1379} (\bibinfo {year} {2015})}\BibitemShut {NoStop}%
\bibitem [{\citenamefont {Dell'Anno}\ \emph {et~al.}(2006)\citenamefont
  {Dell'Anno}, \citenamefont {Siena},\ and\ \citenamefont
  {Illuminati}}]{dellanno06}%
  \BibitemOpen
  \bibfield  {author} {\bibinfo {author} {\bibfnamefont {F.}~\bibnamefont
  {Dell'Anno}}, \bibinfo {author} {\bibfnamefont {S.~D.}\ \bibnamefont
  {Siena}}, \ and\ \bibinfo {author} {\bibfnamefont {F.}~\bibnamefont
  {Illuminati}},\ }\href {\doibase 10.1016/j.physrep.2006.01.004} {\bibfield
  {journal} {\bibinfo  {journal} {Phys. Rep.}\ }\textbf {\bibinfo {volume}
  {428}},\ \bibinfo {pages} {53} (\bibinfo {year} {2006})}\BibitemShut
  {NoStop}%
\bibitem [{\citenamefont {Wang}\ \emph {et~al.}(2016)\citenamefont {Wang},
  \citenamefont {Chen}, \citenamefont {Li}, \citenamefont {Huang},
  \citenamefont {Liu}, \citenamefont {Chen}, \citenamefont {Luo}, \citenamefont
  {Su}, \citenamefont {Wu}, \citenamefont {Li}, \citenamefont {Lu},
  \citenamefont {Hu}, \citenamefont {Jiang}, \citenamefont {Peng},
  \citenamefont {Li}, \citenamefont {Liu}, \citenamefont {Chen}, \citenamefont
  {Lu},\ and\ \citenamefont {Pan}}]{wang16a}%
  \BibitemOpen
  \bibfield  {author} {\bibinfo {author} {\bibfnamefont {X.-L.}\ \bibnamefont
  {Wang}}, \bibinfo {author} {\bibfnamefont {L.-K.}\ \bibnamefont {Chen}},
  \bibinfo {author} {\bibfnamefont {W.}~\bibnamefont {Li}}, \bibinfo {author}
  {\bibfnamefont {H.-L.}\ \bibnamefont {Huang}}, \bibinfo {author}
  {\bibfnamefont {C.}~\bibnamefont {Liu}}, \bibinfo {author} {\bibfnamefont
  {C.}~\bibnamefont {Chen}}, \bibinfo {author} {\bibfnamefont {Y.-H.}\
  \bibnamefont {Luo}}, \bibinfo {author} {\bibfnamefont {Z.-E.}\ \bibnamefont
  {Su}}, \bibinfo {author} {\bibfnamefont {D.}~\bibnamefont {Wu}}, \bibinfo
  {author} {\bibfnamefont {Z.-D.}\ \bibnamefont {Li}}, \bibinfo {author}
  {\bibfnamefont {H.}~\bibnamefont {Lu}}, \bibinfo {author} {\bibfnamefont
  {Y.}~\bibnamefont {Hu}}, \bibinfo {author} {\bibfnamefont {X.}~\bibnamefont
  {Jiang}}, \bibinfo {author} {\bibfnamefont {C.-Z.}\ \bibnamefont {Peng}},
  \bibinfo {author} {\bibfnamefont {L.}~\bibnamefont {Li}}, \bibinfo {author}
  {\bibfnamefont {N.-L.}\ \bibnamefont {Liu}}, \bibinfo {author} {\bibfnamefont
  {Y.-A.}\ \bibnamefont {Chen}}, \bibinfo {author} {\bibfnamefont {C.-Y.}\
  \bibnamefont {Lu}}, \ and\ \bibinfo {author} {\bibfnamefont {J.-W.}\
  \bibnamefont {Pan}},\ }\href {\doibase 10.1103/PhysRevLett.117.210502}
  {\bibfield  {journal} {\bibinfo  {journal} {Phys. Rev. Lett.}\ }\textbf
  {\bibinfo {volume} {117}},\ \bibinfo {pages} {210502} (\bibinfo {year}
  {2016})}\BibitemShut {NoStop}%
\bibitem [{\citenamefont {Vetsch}\ \emph {et~al.}(2010)\citenamefont {Vetsch},
  \citenamefont {Reitz}, \citenamefont {Sagu{\'e}}, \citenamefont {Schmidt},
  \citenamefont {Dawkins},\ and\ \citenamefont {Rauschenbeutel}}]{vetsch10a}%
  \BibitemOpen
  \bibfield  {author} {\bibinfo {author} {\bibfnamefont {E.}~\bibnamefont
  {Vetsch}}, \bibinfo {author} {\bibfnamefont {D.}~\bibnamefont {Reitz}},
  \bibinfo {author} {\bibfnamefont {G.}~\bibnamefont {Sagu{\'e}}}, \bibinfo
  {author} {\bibfnamefont {R.}~\bibnamefont {Schmidt}}, \bibinfo {author}
  {\bibfnamefont {S.}~\bibnamefont {Dawkins}}, \ and\ \bibinfo {author}
  {\bibfnamefont {A.}~\bibnamefont {Rauschenbeutel}},\ }\href@noop {}
  {\bibfield  {journal} {\bibinfo  {journal} {Phys. Rev. Lett.}\ }\textbf
  {\bibinfo {volume} {104}},\ \bibinfo {pages} {203603} (\bibinfo {year}
  {2010})}\BibitemShut {NoStop}%
\bibitem [{\citenamefont {Goban}\ \emph {et~al.}(2014)\citenamefont {Goban},
  \citenamefont {Hung}, \citenamefont {Yu}, \citenamefont {Hood}, \citenamefont
  {Muniz}, \citenamefont {Lee}, \citenamefont {Martin}, \citenamefont
  {McClung}, \citenamefont {Choi}, \citenamefont {Chang}, \citenamefont
  {Painter},\ and\ \citenamefont {Kimble}}]{goban13a}%
  \BibitemOpen
  \bibfield  {author} {\bibinfo {author} {\bibfnamefont {A.}~\bibnamefont
  {Goban}}, \bibinfo {author} {\bibfnamefont {C.-L.}\ \bibnamefont {Hung}},
  \bibinfo {author} {\bibfnamefont {S.-P.}\ \bibnamefont {Yu}}, \bibinfo
  {author} {\bibfnamefont {J.}~\bibnamefont {Hood}}, \bibinfo {author}
  {\bibfnamefont {J.}~\bibnamefont {Muniz}}, \bibinfo {author} {\bibfnamefont
  {J.}~\bibnamefont {Lee}}, \bibinfo {author} {\bibfnamefont {M.}~\bibnamefont
  {Martin}}, \bibinfo {author} {\bibfnamefont {A.}~\bibnamefont {McClung}},
  \bibinfo {author} {\bibfnamefont {K.}~\bibnamefont {Choi}}, \bibinfo {author}
  {\bibfnamefont {D.}~\bibnamefont {Chang}}, \bibinfo {author} {\bibfnamefont
  {O.}~\bibnamefont {Painter}}, \ and\ \bibinfo {author} {\bibfnamefont
  {H.}~\bibnamefont {Kimble}},\ }\href@noop {} {\bibfield  {journal} {\bibinfo
  {journal} {Nat. Commun.}\ }\textbf {\bibinfo {volume} {5}},\ \bibinfo {pages}
  {3808} (\bibinfo {year} {2014})}\BibitemShut {NoStop}%
\bibitem [{\citenamefont {B\'eguin}\ \emph {et~al.}(2014)\citenamefont
  {B\'eguin}, \citenamefont {Bookjans}, \citenamefont {Christensen},
  \citenamefont {S\o{}rensen}, \citenamefont {M\"uller}, \citenamefont
  {Polzik},\ and\ \citenamefont {Appel}}]{beguin14a}%
  \BibitemOpen
  \bibfield  {author} {\bibinfo {author} {\bibfnamefont {J.-B.}\ \bibnamefont
  {B\'eguin}}, \bibinfo {author} {\bibfnamefont {E.~M.}\ \bibnamefont
  {Bookjans}}, \bibinfo {author} {\bibfnamefont {S.~L.}\ \bibnamefont
  {Christensen}}, \bibinfo {author} {\bibfnamefont {H.~L.}\ \bibnamefont
  {S\o{}rensen}}, \bibinfo {author} {\bibfnamefont {J.~H.}\ \bibnamefont
  {M\"uller}}, \bibinfo {author} {\bibfnamefont {E.~S.}\ \bibnamefont
  {Polzik}}, \ and\ \bibinfo {author} {\bibfnamefont {J.}~\bibnamefont
  {Appel}},\ }\href {\doibase 10.1103/PhysRevLett.113.263603} {\bibfield
  {journal} {\bibinfo  {journal} {Phys. Rev. Lett.}\ }\textbf {\bibinfo
  {volume} {113}},\ \bibinfo {pages} {263603} (\bibinfo {year}
  {2014})}\BibitemShut {NoStop}%
\bibitem [{\citenamefont {Goban}\ \emph {et~al.}(2015)\citenamefont {Goban},
  \citenamefont {Hung}, \citenamefont {Hood}, \citenamefont {Yu}, \citenamefont
  {Muniz}, \citenamefont {Painter},\ and\ \citenamefont {Kimble}}]{goban15a}%
  \BibitemOpen
  \bibfield  {author} {\bibinfo {author} {\bibfnamefont {A.}~\bibnamefont
  {Goban}}, \bibinfo {author} {\bibfnamefont {C.-L.}\ \bibnamefont {Hung}},
  \bibinfo {author} {\bibfnamefont {J.~D.}\ \bibnamefont {Hood}}, \bibinfo
  {author} {\bibfnamefont {S.-P.}\ \bibnamefont {Yu}}, \bibinfo {author}
  {\bibfnamefont {J.~A.}\ \bibnamefont {Muniz}}, \bibinfo {author}
  {\bibfnamefont {O.}~\bibnamefont {Painter}}, \ and\ \bibinfo {author}
  {\bibfnamefont {H.~J.}\ \bibnamefont {Kimble}},\ }\href {\doibase
  10.1103/PhysRevLett.115.063601} {\bibfield  {journal} {\bibinfo  {journal}
  {Phys. Rev. Lett.}\ }\textbf {\bibinfo {volume} {115}},\ \bibinfo {pages}
  {063601} (\bibinfo {year} {2015})}\BibitemShut {NoStop}%
\bibitem [{\citenamefont {S\o{}rensen}\ \emph {et~al.}(2016)\citenamefont
  {S\o{}rensen}, \citenamefont {B\'eguin}, \citenamefont {Kluge}, \citenamefont
  {Iakoupov}, \citenamefont {S\o{}rensen}, \citenamefont {M\"uller},
  \citenamefont {Polzik},\ and\ \citenamefont {Appel}}]{sorensen16a}%
  \BibitemOpen
  \bibfield  {author} {\bibinfo {author} {\bibfnamefont {H.~L.}\ \bibnamefont
  {S\o{}rensen}}, \bibinfo {author} {\bibfnamefont {J.-B.}\ \bibnamefont
  {B\'eguin}}, \bibinfo {author} {\bibfnamefont {K.~W.}\ \bibnamefont {Kluge}},
  \bibinfo {author} {\bibfnamefont {I.}~\bibnamefont {Iakoupov}}, \bibinfo
  {author} {\bibfnamefont {A.~S.}\ \bibnamefont {S\o{}rensen}}, \bibinfo
  {author} {\bibfnamefont {J.~H.}\ \bibnamefont {M\"uller}}, \bibinfo {author}
  {\bibfnamefont {E.~S.}\ \bibnamefont {Polzik}}, \ and\ \bibinfo {author}
  {\bibfnamefont {J.}~\bibnamefont {Appel}},\ }\href {\doibase
  10.1103/PhysRevLett.117.133604} {\bibfield  {journal} {\bibinfo  {journal}
  {Phys. Rev. Lett.}\ }\textbf {\bibinfo {volume} {117}},\ \bibinfo {pages}
  {133604} (\bibinfo {year} {2016})}\BibitemShut {NoStop}%
\bibitem [{\citenamefont {Corzo}\ \emph {et~al.}(2016)\citenamefont {Corzo},
  \citenamefont {Gouraud}, \citenamefont {Chandra}, \citenamefont {Goban},
  \citenamefont {Sheremet}, \citenamefont {Kupriyanov},\ and\ \citenamefont
  {Laurat}}]{corzo16a}%
  \BibitemOpen
  \bibfield  {author} {\bibinfo {author} {\bibfnamefont {N.~V.}\ \bibnamefont
  {Corzo}}, \bibinfo {author} {\bibfnamefont {B.}~\bibnamefont {Gouraud}},
  \bibinfo {author} {\bibfnamefont {A.}~\bibnamefont {Chandra}}, \bibinfo
  {author} {\bibfnamefont {A.}~\bibnamefont {Goban}}, \bibinfo {author}
  {\bibfnamefont {A.~S.}\ \bibnamefont {Sheremet}}, \bibinfo {author}
  {\bibfnamefont {D.~V.}\ \bibnamefont {Kupriyanov}}, \ and\ \bibinfo {author}
  {\bibfnamefont {J.}~\bibnamefont {Laurat}},\ }\href {\doibase
  10.1103/PhysRevLett.117.133603} {\bibfield  {journal} {\bibinfo  {journal}
  {Phys. Rev. Lett.}\ }\textbf {\bibinfo {volume} {117}},\ \bibinfo {pages}
  {133603} (\bibinfo {year} {2016})}\BibitemShut {NoStop}%
\bibitem [{\citenamefont {Solano}\ \emph {et~al.}(2017)\citenamefont {Solano},
  \citenamefont {Grover}, \citenamefont {Hoffman}, \citenamefont {Ravets},
  \citenamefont {Fatemi}, \citenamefont {Orozco},\ and\ \citenamefont
  {Rolston}}]{solano17}%
  \BibitemOpen
  \bibfield  {author} {\bibinfo {author} {\bibfnamefont {P.}~\bibnamefont
  {Solano}}, \bibinfo {author} {\bibfnamefont {J.~A.}\ \bibnamefont {Grover}},
  \bibinfo {author} {\bibfnamefont {J.~E.}\ \bibnamefont {Hoffman}}, \bibinfo
  {author} {\bibfnamefont {S.}~\bibnamefont {Ravets}}, \bibinfo {author}
  {\bibfnamefont {F.~K.}\ \bibnamefont {Fatemi}}, \bibinfo {author}
  {\bibfnamefont {L.~A.}\ \bibnamefont {Orozco}}, \ and\ \bibinfo {author}
  {\bibfnamefont {S.~L.}\ \bibnamefont {Rolston}},\ }in\ \href {\doibase
  10.1016/bs.aamop.2017.02.003} {\emph {\bibinfo {booktitle} {Advances In
  Atomic, Molecular, and Optical Physics}}}\ (\bibinfo  {publisher}
  {Elsevier},\ \bibinfo {year} {2017})\ pp.\ \bibinfo {pages}
  {439--505}\BibitemShut {NoStop}%
\bibitem [{\citenamefont {Lodahl}\ \emph {et~al.}(2015)\citenamefont {Lodahl},
  \citenamefont {Mahmoodian},\ and\ \citenamefont {Stobbe}}]{lodahl15}%
  \BibitemOpen
  \bibfield  {author} {\bibinfo {author} {\bibfnamefont {P.}~\bibnamefont
  {Lodahl}}, \bibinfo {author} {\bibfnamefont {S.}~\bibnamefont {Mahmoodian}},
  \ and\ \bibinfo {author} {\bibfnamefont {S.}~\bibnamefont {Stobbe}},\ }\href
  {\doibase 10.1103/revmodphys.87.347} {\bibfield  {journal} {\bibinfo
  {journal} {Rev. Mod. Phys.}\ }\textbf {\bibinfo {volume} {87}},\ \bibinfo
  {pages} {347} (\bibinfo {year} {2015})}\BibitemShut {NoStop}%
\bibitem [{\citenamefont {{Le Kien}}\ \emph {et~al.}(2005)\citenamefont {{Le
  Kien}}, \citenamefont {Gupta}, \citenamefont {Nayak},\ and\ \citenamefont
  {Hakuta}}]{kien05a}%
  \BibitemOpen
  \bibfield  {author} {\bibinfo {author} {\bibfnamefont {F.}~\bibnamefont {{Le
  Kien}}}, \bibinfo {author} {\bibfnamefont {S.~D.}\ \bibnamefont {Gupta}},
  \bibinfo {author} {\bibfnamefont {K.~P.}\ \bibnamefont {Nayak}}, \ and\
  \bibinfo {author} {\bibfnamefont {K.}~\bibnamefont {Hakuta}},\ }\href
  {\doibase 10.1103/PhysRevA.72.063815} {\bibfield  {journal} {\bibinfo
  {journal} {Phys. Rev. A}\ }\textbf {\bibinfo {volume} {72}},\ \bibinfo
  {pages} {063815} (\bibinfo {year} {2005})}\BibitemShut {NoStop}%
\bibitem [{\citenamefont {Chang}\ \emph {et~al.}(2012)\citenamefont {Chang},
  \citenamefont {Jiang}, \citenamefont {Gorshkov},\ and\ \citenamefont
  {Kimble}}]{chang12a}%
  \BibitemOpen
  \bibfield  {author} {\bibinfo {author} {\bibfnamefont {D.}~\bibnamefont
  {Chang}}, \bibinfo {author} {\bibfnamefont {L.}~\bibnamefont {Jiang}},
  \bibinfo {author} {\bibfnamefont {A.}~\bibnamefont {Gorshkov}}, \ and\
  \bibinfo {author} {\bibfnamefont {H.}~\bibnamefont {Kimble}},\ }\href@noop {}
  {\bibfield  {journal} {\bibinfo  {journal} {New J. Phys.}\ }\textbf {\bibinfo
  {volume} {14}},\ \bibinfo {pages} {063003} (\bibinfo {year}
  {2012})}\BibitemShut {NoStop}%
\bibitem [{\citenamefont {Porras}\ and\ \citenamefont
  {Cirac}(2008)}]{porras08a}%
  \BibitemOpen
  \bibfield  {author} {\bibinfo {author} {\bibfnamefont {D.}~\bibnamefont
  {Porras}}\ and\ \bibinfo {author} {\bibfnamefont {J.}~\bibnamefont {Cirac}},\
  }\href@noop {} {\bibfield  {journal} {\bibinfo  {journal} {Phys. Rev. A}\
  }\textbf {\bibinfo {volume} {78}},\ \bibinfo {pages} {053816} (\bibinfo
  {year} {2008})}\BibitemShut {NoStop}%
\bibitem [{\citenamefont {Gonz\'alez-Tudela}\ \emph {et~al.}(2015)\citenamefont
  {Gonz\'alez-Tudela}, \citenamefont {Paulisch}, \citenamefont {Chang},
  \citenamefont {Kimble},\ and\ \citenamefont {Cirac}}]{gonzaleztudela15a}%
  \BibitemOpen
  \bibfield  {author} {\bibinfo {author} {\bibfnamefont {A.}~\bibnamefont
  {Gonz\'alez-Tudela}}, \bibinfo {author} {\bibfnamefont {V.}~\bibnamefont
  {Paulisch}}, \bibinfo {author} {\bibfnamefont {D.~E.}\ \bibnamefont {Chang}},
  \bibinfo {author} {\bibfnamefont {H.~J.}\ \bibnamefont {Kimble}}, \ and\
  \bibinfo {author} {\bibfnamefont {J.~I.}\ \bibnamefont {Cirac}},\ }\href
  {\doibase 10.1103/PhysRevLett.115.163603} {\bibfield  {journal} {\bibinfo
  {journal} {Phys. Rev. Lett.}\ }\textbf {\bibinfo {volume} {115}},\ \bibinfo
  {pages} {163603} (\bibinfo {year} {2015})}\BibitemShut {NoStop}%
\bibitem [{\citenamefont {Fiur\'a\ifmmode~\check{s}\else \v{s}\fi{}ek}\ \emph
  {et~al.}(2005)\citenamefont {Fiur\'a\ifmmode~\check{s}\else \v{s}\fi{}ek},
  \citenamefont {Garc\'{i}a-Patr\'on},\ and\ \citenamefont
  {Cerf}}]{fiurasek05a}%
  \BibitemOpen
  \bibfield  {author} {\bibinfo {author} {\bibfnamefont {J.}~\bibnamefont
  {Fiur\'a\ifmmode~\check{s}\else \v{s}\fi{}ek}}, \bibinfo {author}
  {\bibfnamefont {R.}~\bibnamefont {Garc\'{i}a-Patr\'on}}, \ and\ \bibinfo
  {author} {\bibfnamefont {N.~J.}\ \bibnamefont {Cerf}},\ }\href {\doibase
  10.1103/PhysRevA.72.033822} {\bibfield  {journal} {\bibinfo  {journal} {Phys.
  Rev. A}\ }\textbf {\bibinfo {volume} {72}},\ \bibinfo {pages} {033822}
  (\bibinfo {year} {2005})}\BibitemShut {NoStop}%
\bibitem [{\citenamefont {Yurke}(1986)}]{yurke86}%
  \BibitemOpen
  \bibfield  {author} {\bibinfo {author} {\bibfnamefont {B.}~\bibnamefont
  {Yurke}},\ }\href {\doibase 10.1103/physrevlett.56.1515} {\bibfield
  {journal} {\bibinfo  {journal} {Phys. Rev. Lett.}\ }\textbf {\bibinfo
  {volume} {56}},\ \bibinfo {pages} {1515} (\bibinfo {year}
  {1986})}\BibitemShut {NoStop}%
\bibitem [{\citenamefont {Gardiner}\ and\ \citenamefont
  {Zoller}(2000)}]{gardiner_book00a}%
  \BibitemOpen
  \bibfield  {author} {\bibinfo {author} {\bibfnamefont {G.~W.}\ \bibnamefont
  {Gardiner}}\ and\ \bibinfo {author} {\bibfnamefont {P.}~\bibnamefont
  {Zoller}},\ }\href@noop {} {\emph {\bibinfo {title} {{Quantum Noise}}}},\
  \bibinfo {edition} {2nd}\ ed.\ (\bibinfo  {publisher} {Springer-Verlag,
  Berlin},\ \bibinfo {year} {2000})\BibitemShut {NoStop}%
\bibitem [{\citenamefont {Asenjo-Garcia}\ \emph {et~al.}(2017)\citenamefont
  {Asenjo-Garcia}, \citenamefont {Moreno-Cardoner}, \citenamefont {Albrecht},
  \citenamefont {Kimble},\ and\ \citenamefont {Chang}}]{asenjogarcia17a}%
  \BibitemOpen
  \bibfield  {author} {\bibinfo {author} {\bibfnamefont {A.}~\bibnamefont
  {Asenjo-Garcia}}, \bibinfo {author} {\bibfnamefont {M.}~\bibnamefont
  {Moreno-Cardoner}}, \bibinfo {author} {\bibfnamefont {A.}~\bibnamefont
  {Albrecht}}, \bibinfo {author} {\bibfnamefont {H.~J.}\ \bibnamefont
  {Kimble}}, \ and\ \bibinfo {author} {\bibfnamefont {D.~E.}\ \bibnamefont
  {Chang}},\ }\href {\doibase 10.1103/PhysRevX.7.031024} {\bibfield  {journal}
  {\bibinfo  {journal} {Phys. Rev. X}\ }\textbf {\bibinfo {volume} {7}},\
  \bibinfo {pages} {031024} (\bibinfo {year} {2017})}\BibitemShut {NoStop}%
\bibitem [{\citenamefont {Hughes}\ and\ \citenamefont {Yao}(2009)}]{hughes09a}%
  \BibitemOpen
  \bibfield  {author} {\bibinfo {author} {\bibfnamefont {S.}~\bibnamefont
  {Hughes}}\ and\ \bibinfo {author} {\bibfnamefont {P.}~\bibnamefont {Yao}},\
  }\href {\doibase 10.1364/OE.17.003322} {\bibfield  {journal} {\bibinfo
  {journal} {Opt. Express}\ }\textbf {\bibinfo {volume} {17}},\ \bibinfo
  {pages} {3322} (\bibinfo {year} {2009})}\BibitemShut {NoStop}%
\bibitem [{\citenamefont {Laucht}\ \emph
  {et~al.}(2012{\natexlab{b}})\citenamefont {Laucht}, \citenamefont {Pütz},
  \citenamefont {Günthner}, \citenamefont {Hauke}, \citenamefont {Saive},
  \citenamefont {Fr{\'{e}}d{\'{e}}rick}, \citenamefont {Bichler}, \citenamefont
  {Amann}, \citenamefont {Holleitner}, \citenamefont {Kaniber},\ and\
  \citenamefont {Finley}}]{laucht12}%
  \BibitemOpen
  \bibfield  {author} {\bibinfo {author} {\bibfnamefont {A.}~\bibnamefont
  {Laucht}}, \bibinfo {author} {\bibfnamefont {S.}~\bibnamefont {Pütz}},
  \bibinfo {author} {\bibfnamefont {T.}~\bibnamefont {Günthner}}, \bibinfo
  {author} {\bibfnamefont {N.}~\bibnamefont {Hauke}}, \bibinfo {author}
  {\bibfnamefont {R.}~\bibnamefont {Saive}}, \bibinfo {author} {\bibfnamefont
  {S.}~\bibnamefont {Fr{\'{e}}d{\'{e}}rick}}, \bibinfo {author} {\bibfnamefont
  {M.}~\bibnamefont {Bichler}}, \bibinfo {author} {\bibfnamefont {M.-C.}\
  \bibnamefont {Amann}}, \bibinfo {author} {\bibfnamefont {A.~W.}\ \bibnamefont
  {Holleitner}}, \bibinfo {author} {\bibfnamefont {M.}~\bibnamefont {Kaniber}},
  \ and\ \bibinfo {author} {\bibfnamefont {J.~J.}\ \bibnamefont {Finley}},\
  }\href {\doibase 10.1103/physrevx.2.011014} {\bibfield  {journal} {\bibinfo
  {journal} {Phys. Rev. X}\ }\textbf {\bibinfo {volume} {2}} (\bibinfo {year}
  {2012}{\natexlab{b}}),\ 10.1103/physrevx.2.011014}\BibitemShut {NoStop}%
\bibitem [{\citenamefont {Yu}\ \emph {et~al.}(2014)\citenamefont {Yu},
  \citenamefont {Hood}, \citenamefont {Muniz}, \citenamefont {Martin},
  \citenamefont {Norte}, \citenamefont {Hung}, \citenamefont {Meenehan},
  \citenamefont {Cohen}, \citenamefont {Painter},\ and\ \citenamefont
  {Kimble}}]{yu14a}%
  \BibitemOpen
  \bibfield  {author} {\bibinfo {author} {\bibfnamefont {S.-P.}\ \bibnamefont
  {Yu}}, \bibinfo {author} {\bibfnamefont {J.}~\bibnamefont {Hood}}, \bibinfo
  {author} {\bibfnamefont {J.}~\bibnamefont {Muniz}}, \bibinfo {author}
  {\bibfnamefont {M.}~\bibnamefont {Martin}}, \bibinfo {author} {\bibfnamefont
  {R.}~\bibnamefont {Norte}}, \bibinfo {author} {\bibfnamefont {C.-L.}\
  \bibnamefont {Hung}}, \bibinfo {author} {\bibfnamefont {S.~M.}\ \bibnamefont
  {Meenehan}}, \bibinfo {author} {\bibfnamefont {J.~D.}\ \bibnamefont {Cohen}},
  \bibinfo {author} {\bibfnamefont {O.}~\bibnamefont {Painter}}, \ and\
  \bibinfo {author} {\bibfnamefont {H.}~\bibnamefont {Kimble}},\ }\href@noop {}
  {\bibfield  {journal} {\bibinfo  {journal} {Appl. Phys. Lett.}\ }\textbf
  {\bibinfo {volume} {104}},\ \bibinfo {pages} {111103} (\bibinfo {year}
  {2014})}\BibitemShut {NoStop}%
\bibitem [{\citenamefont {van Enk}\ \emph {et~al.}(1997)\citenamefont {van
  Enk}, \citenamefont {Cirac},\ and\ \citenamefont {Zoller}}]{enk97}%
  \BibitemOpen
  \bibfield  {author} {\bibinfo {author} {\bibfnamefont {S.~J.}\ \bibnamefont
  {van Enk}}, \bibinfo {author} {\bibfnamefont {J.~I.}\ \bibnamefont {Cirac}},
  \ and\ \bibinfo {author} {\bibfnamefont {P.}~\bibnamefont {Zoller}},\ }\href
  {\doibase 10.1103/physrevlett.78.4293} {\bibfield  {journal} {\bibinfo
  {journal} {Phys. Rev. Lett.}\ }\textbf {\bibinfo {volume} {78}},\ \bibinfo
  {pages} {4293} (\bibinfo {year} {1997})}\BibitemShut {NoStop}%
\bibitem [{\citenamefont {Borregaard}\ \emph {et~al.}(2015)\citenamefont
  {Borregaard}, \citenamefont {K\'om\'ar}, \citenamefont {Kessler},
  \citenamefont {S\o{}rensen},\ and\ \citenamefont {Lukin}}]{borregaard15a}%
  \BibitemOpen
  \bibfield  {author} {\bibinfo {author} {\bibfnamefont {J.}~\bibnamefont
  {Borregaard}}, \bibinfo {author} {\bibfnamefont {P.}~\bibnamefont
  {K\'om\'ar}}, \bibinfo {author} {\bibfnamefont {E.~M.}\ \bibnamefont
  {Kessler}}, \bibinfo {author} {\bibfnamefont {A.~S.}\ \bibnamefont
  {S\o{}rensen}}, \ and\ \bibinfo {author} {\bibfnamefont {M.~D.}\ \bibnamefont
  {Lukin}},\ }\href {\doibase 10.1103/PhysRevLett.114.110502} {\bibfield
  {journal} {\bibinfo  {journal} {Phys. Rev. Lett.}\ }\textbf {\bibinfo
  {volume} {114}},\ \bibinfo {pages} {110502} (\bibinfo {year}
  {2015})}\BibitemShut {NoStop}%
\bibitem [{\citenamefont {Facchi}\ and\ \citenamefont
  {Pascazio}(2002)}]{facchi02a}%
  \BibitemOpen
  \bibfield  {author} {\bibinfo {author} {\bibfnamefont {P.}~\bibnamefont
  {Facchi}}\ and\ \bibinfo {author} {\bibfnamefont {S.}~\bibnamefont
  {Pascazio}},\ }\href {\doibase 10.1103/PhysRevLett.89.080401} {\bibfield
  {journal} {\bibinfo  {journal} {Phys. Rev. Lett.}\ }\textbf {\bibinfo
  {volume} {89}},\ \bibinfo {pages} {080401} (\bibinfo {year}
  {2002})}\BibitemShut {NoStop}%
\bibitem [{\citenamefont {Holstein}\ and\ \citenamefont
  {Primakoff}(1940{\natexlab{a}})}]{Holstein1940}%
  \BibitemOpen
  \bibfield  {author} {\bibinfo {author} {\bibfnamefont {T.}~\bibnamefont
  {Holstein}}\ and\ \bibinfo {author} {\bibfnamefont {H.}~\bibnamefont
  {Primakoff}},\ }\href {\doibase 10.1103/PhysRev.58.1098} {\bibfield
  {journal} {\bibinfo  {journal} {Phys. Rev.}\ }\textbf {\bibinfo {volume}
  {58}},\ \bibinfo {pages} {1098} (\bibinfo {year}
  {1940}{\natexlab{a}})}\BibitemShut {NoStop}%
\bibitem [{\citenamefont {Motes}\ \emph {et~al.}(2016)\citenamefont {Motes},
  \citenamefont {Mann}, \citenamefont {Olson}, \citenamefont {Studer},
  \citenamefont {Bergeron}, \citenamefont {Gilchrist}, \citenamefont {Dowling},
  \citenamefont {Berry},\ and\ \citenamefont {Rohde}}]{motes16}%
  \BibitemOpen
  \bibfield  {author} {\bibinfo {author} {\bibfnamefont {K.~R.}\ \bibnamefont
  {Motes}}, \bibinfo {author} {\bibfnamefont {R.~L.}\ \bibnamefont {Mann}},
  \bibinfo {author} {\bibfnamefont {J.~P.}\ \bibnamefont {Olson}}, \bibinfo
  {author} {\bibfnamefont {N.~M.}\ \bibnamefont {Studer}}, \bibinfo {author}
  {\bibfnamefont {E.~A.}\ \bibnamefont {Bergeron}}, \bibinfo {author}
  {\bibfnamefont {A.}~\bibnamefont {Gilchrist}}, \bibinfo {author}
  {\bibfnamefont {J.~P.}\ \bibnamefont {Dowling}}, \bibinfo {author}
  {\bibfnamefont {D.~W.}\ \bibnamefont {Berry}}, \ and\ \bibinfo {author}
  {\bibfnamefont {P.~P.}\ \bibnamefont {Rohde}},\ }\href {\doibase
  10.1103/PhysRevA.94.012344} {\bibfield  {journal} {\bibinfo  {journal} {Phys.
  Rev. A}\ }\textbf {\bibinfo {volume} {94}},\ \bibinfo {pages} {012344}
  (\bibinfo {year} {2016})},\ \Eprint {http://arxiv.org/abs/1603.00533}
  {1603.00533} \BibitemShut {NoStop}%
\bibitem [{\citenamefont {Holstein}\ and\ \citenamefont
  {Primakoff}(1940{\natexlab{b}})}]{holstein40a}%
  \BibitemOpen
  \bibfield  {author} {\bibinfo {author} {\bibfnamefont {T.}~\bibnamefont
  {Holstein}}\ and\ \bibinfo {author} {\bibfnamefont {H.}~\bibnamefont
  {Primakoff}},\ }\href {\doibase 10.1103/PhysRev.58.1098} {\bibfield
  {journal} {\bibinfo  {journal} {Phys. Rev.}\ }\textbf {\bibinfo {volume}
  {58}},\ \bibinfo {pages} {1098} (\bibinfo {year}
  {1940}{\natexlab{b}})}\BibitemShut {NoStop}%
\bibitem [{\citenamefont {Demkowicz-Dobrza{\'{n}}ski}\ \emph
  {et~al.}(2015)\citenamefont {Demkowicz-Dobrza{\'{n}}ski}, \citenamefont
  {Jarzyna},\ and\ \citenamefont
  {Ko{\l}ody{\'{n}}ski}}]{demkowicz-dobrzanski15}%
  \BibitemOpen
  \bibfield  {author} {\bibinfo {author} {\bibfnamefont {R.}~\bibnamefont
  {Demkowicz-Dobrza{\'{n}}ski}}, \bibinfo {author} {\bibfnamefont
  {M.}~\bibnamefont {Jarzyna}}, \ and\ \bibinfo {author} {\bibfnamefont
  {J.}~\bibnamefont {Ko{\l}ody{\'{n}}ski}},\ }in\ \href {\doibase
  10.1016/bs.po.2015.02.003} {\emph {\bibinfo {booktitle} {Progress in
  Optics}}}\ (\bibinfo  {publisher} {Elsevier},\ \bibinfo {year} {2015})\ pp.\
  \bibinfo {pages} {345--435}\BibitemShut {NoStop}%
\bibitem [{\citenamefont {Campos}\ \emph {et~al.}(2003)\citenamefont {Campos},
  \citenamefont {Gerry},\ and\ \citenamefont {Benmoussa}}]{campos03}%
  \BibitemOpen
  \bibfield  {author} {\bibinfo {author} {\bibfnamefont {R.~A.}\ \bibnamefont
  {Campos}}, \bibinfo {author} {\bibfnamefont {C.~C.}\ \bibnamefont {Gerry}}, \
  and\ \bibinfo {author} {\bibfnamefont {A.}~\bibnamefont {Benmoussa}},\ }\href
  {\doibase 10.1103/physreva.68.023810} {\bibfield  {journal} {\bibinfo
  {journal} {Phys. Rev. A}\ }\textbf {\bibinfo {volume} {68}} (\bibinfo {year}
  {2003}),\ 10.1103/physreva.68.023810}\BibitemShut {NoStop}%
\bibitem [{\citenamefont {Cooper}\ \emph {et~al.}(2010)\citenamefont {Cooper},
  \citenamefont {Hallwood},\ and\ \citenamefont {Dunningham}}]{cooper10}%
  \BibitemOpen
  \bibfield  {author} {\bibinfo {author} {\bibfnamefont {J.~J.}\ \bibnamefont
  {Cooper}}, \bibinfo {author} {\bibfnamefont {D.~W.}\ \bibnamefont
  {Hallwood}}, \ and\ \bibinfo {author} {\bibfnamefont {J.~A.}\ \bibnamefont
  {Dunningham}},\ }\href {\doibase 10.1103/physreva.81.043624} {\bibfield
  {journal} {\bibinfo  {journal} {Phys. Rev. A}\ }\textbf {\bibinfo {volume}
  {81}} (\bibinfo {year} {2010}),\ 10.1103/physreva.81.043624}\BibitemShut
  {NoStop}%
\bibitem [{\citenamefont {Ono}\ \emph {et~al.}(2016)\citenamefont {Ono},
  \citenamefont {Chesterking}, \citenamefont {Cable}, \citenamefont {O'Brien},\
  and\ \citenamefont {Matthews}}]{ono16}%
  \BibitemOpen
  \bibfield  {author} {\bibinfo {author} {\bibfnamefont {T.}~\bibnamefont
  {Ono}}, \bibinfo {author} {\bibfnamefont {J.~S.}\ \bibnamefont
  {Chesterking}}, \bibinfo {author} {\bibfnamefont {H.}~\bibnamefont {Cable}},
  \bibinfo {author} {\bibfnamefont {J.~L.}\ \bibnamefont {O'Brien}}, \ and\
  \bibinfo {author} {\bibfnamefont {J.~C.~F.}\ \bibnamefont {Matthews}},\
  }\href {\doibase 10.1088/1367-2630/aa6e39} {\bibfield  {journal} {\bibinfo
  {journal} {New J. Phys.}\ } (\bibinfo {year} {2016}),\
  10.1088/1367-2630/aa6e39},\ \Eprint {http://arxiv.org/abs/1610.03790v1}
  {1610.03790v1} \BibitemShut {NoStop}%
\bibitem [{\citenamefont {Kok}\ \emph {et~al.}(2002)\citenamefont {Kok},
  \citenamefont {Lee},\ and\ \citenamefont {Dowling}}]{kok02}%
  \BibitemOpen
  \bibfield  {author} {\bibinfo {author} {\bibfnamefont {P.}~\bibnamefont
  {Kok}}, \bibinfo {author} {\bibfnamefont {H.}~\bibnamefont {Lee}}, \ and\
  \bibinfo {author} {\bibfnamefont {J.~P.}\ \bibnamefont {Dowling}},\ }\href
  {\doibase 10.1103/physreva.65.052104} {\bibfield  {journal} {\bibinfo
  {journal} {Phys. Rev. A}\ }\textbf {\bibinfo {volume} {65}} (\bibinfo {year}
  {2002}),\ 10.1103/physreva.65.052104}\BibitemShut {NoStop}%
\bibitem [{\citenamefont {Paulisch}\ \emph {et~al.}(2017)\citenamefont
  {Paulisch}, \citenamefont {Gonz\'{a}lez-Tudela}, \citenamefont {Kimble},\
  and\ \citenamefont {Cirac}}]{paulisch17a}%
  \BibitemOpen
  \bibfield  {author} {\bibinfo {author} {\bibfnamefont {V.}~\bibnamefont
  {Paulisch}}, \bibinfo {author} {\bibfnamefont {A.}~\bibnamefont
  {Gonz\'{a}lez-Tudela}}, \bibinfo {author} {\bibfnamefont {H.~J.}\
  \bibnamefont {Kimble}}, \ and\ \bibinfo {author} {\bibfnamefont {J.~I.}\
  \bibnamefont {Cirac}},\ }\href@noop {} {\bibfield  {journal} {\bibinfo
  {journal} {New J. Phys.}\ }\textbf {\bibinfo {volume} {19}},\ \bibinfo
  {pages} {043004} (\bibinfo {year} {2017})}\BibitemShut {NoStop}%
\end{thebibliography}%

\appendix
\section{General Quantum Zeno step for protocol 1 and 2}\label{appendix:zeno}

In protocol 1 and 2 we apply a Quantum Zeno step for the transitions 
\begin{align}
	\ket{e_1}_\rs \ket{\phi_m^{\rs_1}}_\rt \ket{s_1}^{\otimes N_\rd}_\rd
	&\rightarrow
	\ket{g}_\rs S_{cg}^{(\rt)}\ket{\phi_m^{\rs_1}}_\rt \ket{s_1}^{\otimes N_\rd}_\rd\,,\\
	\ket{c}_\rs S_{e_2 g}^{(\rt)}\ket{\phi_m^{\rs}}_\rt \ket{s}^{\otimes N_\rd}_\rd
	&\rightarrow
	\ket{c}_\rs \ket{\phi_{m+1}^{\rs}}_\rt S_{cs}^{(\rd)} \ket{s}^{\otimes N_\rd}_\rd.
\end{align}
In this Section we discuss a generalized Quantum Zeno step and derive the probability of a successful transfer and errors due to spontaneous emission events. We only need to treat two ensembles as either the detector ensemble or the source atom is decoupled from the dynamics.

\begin{figure}[t]
	\centering
	\includegraphics[width=0.45\textwidth]{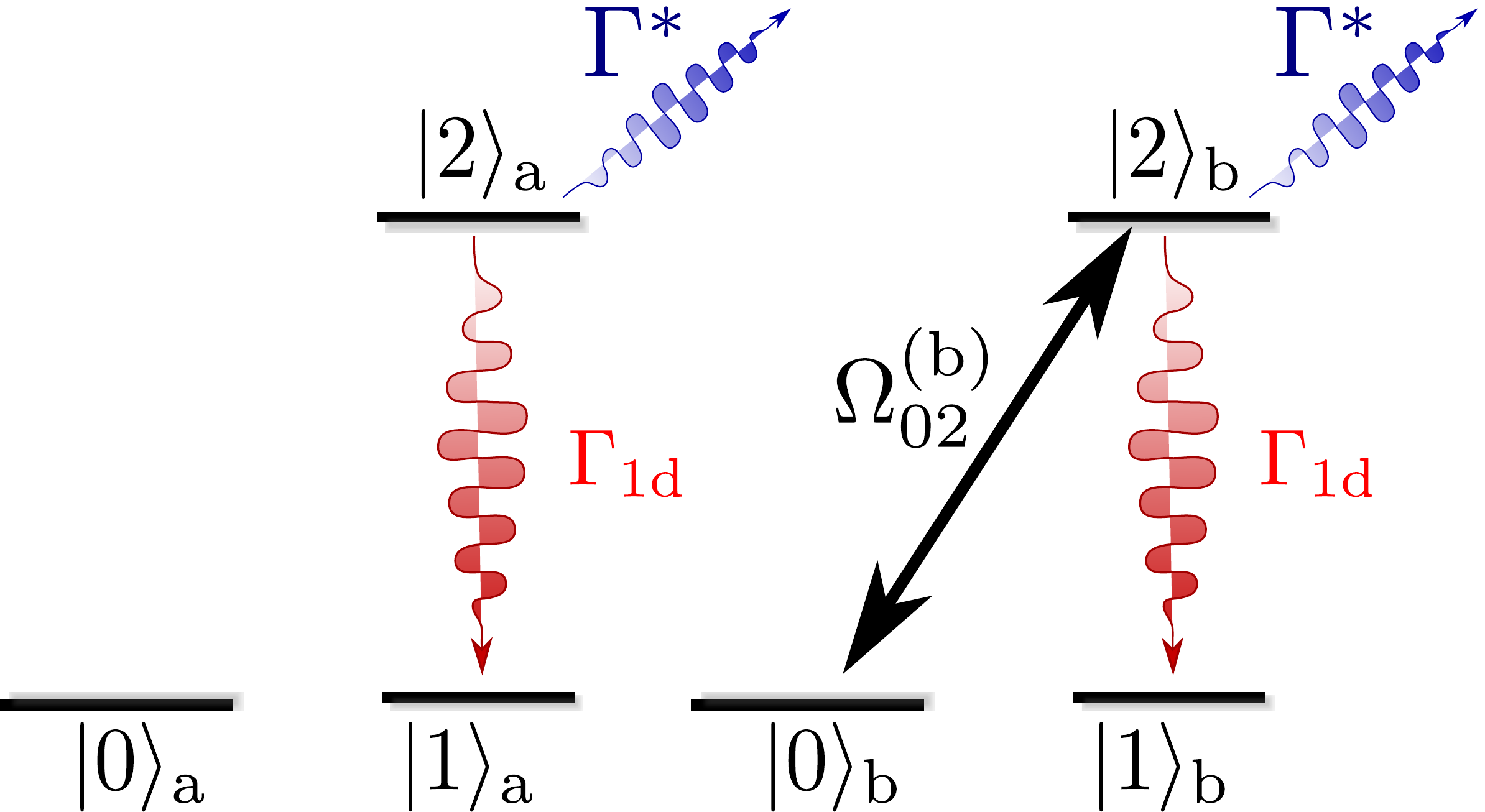}
	\caption{ For the generalized Zeno Step we consider two ensembles with three-level atoms with one transition coupled to a waveguide mode. }
	
	\label{fig:Zeno_General}
\end{figure}

\subsection{System}

Let us assume we have two ensembles (a and b) with three-level atoms (with metastable states $\ket{0}$, $\ket{1}$ and excited state $\ket{2}$) that contain $N_\aa$ and $N_\bb$ atoms each. The dynamics are governed by the collective decay on the $1 \leftrightarrow 2$ transition and an external field with Rabi coupling $\Omega_{02}^{(\bb)}$ on the second ensemble is applied (see Figure \ref{fig:Zeno_General}). Therefore, the effective non-hermitian Hamiltonian that drives the no-jump evolution is given by
\begin{align}
H_\mathrm{eff} =& 
	\frac{1}{2} \Omega_{02}^{(\bb)}  \left( S_{20}^{(\bb)} + S_{02}^{(\bb)} \right)  -\mathrm{i} \frac{\Gamma^*}{2}\sum_{n}\sigma^{n}_{ee}
	\nonumber \\
	&-\mathrm{i} \frac{\Gamma_\oned}{2}
	\big(S_{21}^{(\aa)} + S_{21}^{(\bb)} \big) \big(S_{12}^{(\aa)} + S_{12}^{(\bb)} \big)
	-\mathrm{i} \frac{\Gamma^*}{2}\sum_{n}\sigma^{n}_{ee}. \notag
\end{align}

We denote the collective symmetric excitations as $\ket{\#_0, \#_1, \#_2}_{\aa/\bb} \propto S_{10}^ {\#1} S_{20}^ {\#2} \ket{0}_{\aa/\bb}^{N_{\aa/\bb}}$. The initial state  $\ket{\psi_1}$ is an excited state of ensemble $\aa$ and the state we want to reach, $\ket{\psi_3}$, contains one excitation more in the metastable state $1$. They are coupled via the state $\ket{\psi_2}$, which is also excited, but where the excitation is already in ensemble $\bb$. The states can be written in general as
\begin{align}
\ket{\psi_1}
	 =& \ket{N_\aa - k - 1, k, 1}_\aa \otimes \ket{0, N_\bb, 0} \\
\ket{\psi_2} 
	=& \ket{N_\aa - k - 1, k+1, 0}_\aa \otimes \ket{0, N_\bb-1, 1} \\
\ket{\psi_3} 
	=& \ket{N_\aa - k - 1, k+1, 0}_\aa \otimes \ket{1, N_\bb-1, 0}
\end{align}
In this basis, one can write the non-hermitian Hamiltonian as
\begin{align}
H_\mathrm{eff} 
	=& \frac{1}{2} \Omega_{02}^{(\bb)} 
		\left( \ket{\psi_2} \bra{\psi_3} + \ket{\psi_3} \bra{\psi_2}\right)  
		- \ii \frac{\Gamma^*}{2} \sum_{j = \rs, \rd} \ket{\psi_j} \bra{\psi_j} \notag \\
	& - \ii \frac{\Gamma_\oned}{2} \left(N_\bb + k + 1 \right) \ket{\psi_\rs} \bra{\psi_\rs},
\end{align}
where the decay terms have been expressed in terms of the superradiant and dark state of the combined $\aa$ and $\bb$ ensembles, that are
\begin{align}
\ket{\psi_\rs} 
	=& \sqrt{\frac{k+1}{N_\bb + k+1}} \ket{\psi_1} 
		+ \sqrt{\frac{N_\bb}{N_\bb + k+1}} \ket{\psi_2}, \\
\ket{\psi_\rd} 
	=& \sqrt{\frac{N_b}{N_\bb + k+1}} \ket{\psi_1} 
		- \sqrt{\frac{k+1}{N_\bb + k+1}} \ket{\psi_2}.
\end{align}

\subsection{Non-hermitian evolution: probabilities}

When the coherent driving is weak compared to the collective dissipation $\Omega_{02}^{(\bb)} \ll N_\bb \Gamma_\oned$, the superradiant state can be adiabatically eliminated, which leads to an effective decay rate into the waveguide of the otherwise dark state $\ket{\phi_3}$ with rate  $\frac{N_\bb |\Omega_{02}^{(\bb)}|^2}{(N_\bb + k + 1)^2\Gamma_\oned}$. To minimize the total errors, one chooses $\Omega_{02}^{(\bb)} = \sqrt{(N_\bb + k +1) \Gamma_\oned \Gamma^*}$ such that both dark states have the same decay rate, i.e., $\frac{N_\bb |\Omega_{02}^{(\bb)}|^2}{(N_\bb + k + 1)^2\Gamma_\oned} = \Gamma^*$.

The population in $\ket{\phi_3}$ then follows 
\begin{equation}
|\psi_3(t)|^2
	\approx \frac{N_\bb}{N_\bb + k+1}e^{-\Gamma^*t } \sin^2\Big(t \frac{\Gamma^*\sqrt{(k+1)P_\oned}}{2}\Big)
\end{equation}
where the prefactor originates in the overlap between the initial state and the dark state. Therefore, the success probability $p=|\psi_3(T)|^2$ is maximized for $T = \pi \left( \Gamma^* \sqrt{ (k+1) P_\oned} \right)^{-1}$, yielding
\begin{equation}
p = \frac{N_\bb}{N_\bb + k+1} \ee^{- \pi / \sqrt{(k+1) P_\oned}},
\end{equation}
which can then be particularized to the evolution in the corresponding steps of the protocols.

\subsection{Quantum jump evolution: errors}

To properly analyze the errors of the protocols using Quantum Zeno Dynamics, it is important to know the probability of spontaneous jumps in both ensembles during the evolution.  Typically, the problematic processes are the ones associated to leaky photons. The quantum jump analysis shows that the probability for a spontaneous jump in the ensemble $a$ or $b$ is given by
\begin{align}
p_{\aa,*}&=p_{\aa_1,*}+p_{\aa_2,*}=\Gamma^*\int_0^{T} \rd t |\psi_1(t)|^2\\
	&+ \Gamma^*\int_{0}^\infty \rd t |\langle \psi_\rs | \psi_1 \rangle|^2 e^{-((N_\bb +k + 1) \Gamma_\oned +\Gamma^*) t}\,,  \nonumber \\
p_{\bb,*}&=p_{\bb_1,*}+p_{\bb_2,*}=\Gamma^*\int_0^{T} \rd t |\psi_2(t)|^2 \\
	&+ \Gamma^*\int_{0}^\infty \rd t  |\langle \psi_\rs | \psi_2 \rangle|^2 e^{-((N_\bb +k + 1) \Gamma_\oned +\Gamma^*) t}\,, \nonumber
\end{align}
where the first parts, $p_{\aa/\bb_1,*}$, correspond to the interval of time $(0,T)$ where $\Omega_{02}^\bb$ is switched on, and the second part, $p_{\aa/\bb_2,*}$, comes from population remaining in the excited state after the external field has been switched off and which will decay exponentially until all the population in the excited state, if any, disappears. Here, we have already assumend that there remains no population in the dark state, i.e. $|\psi_\rd(T)|=0$, and the decay only comes from the overlap with the superradiant state. By using the approximations for $H_\mathrm{eff}$, we can calculate the different contributions and upper bound the probabilities from these processes as
\begin{align}
p_{\aa_1,*}
	\lesssim \frac{1}{2}(1-e^{-\pi / \sqrt{(k+1) P_\oned}})
	\lesssim \frac{\pi}{2 \sqrt{P_\oned}}\,\\
p_{\bb_1,*}
	\lesssim \frac{k+1}{2 N_\bb}(1-e^{-\pi / \sqrt{(k+1) P_\oned}})
	\lesssim \frac{\pi \sqrt{k+1}}{2 N_\bb\sqrt{P_\oned}}\,,
\end{align}
which mainly comes from the contribution of the dark state and where the last approximation is valid for $P_\oned\gg 1$. This also shows why step (e) of protocol 2 needs to be modified: When the first ensemble is the target ensemble, the probability of a spontaneous emission event has bad scaling.

Finally, we need to consider what happens with the contribution $p_{\aa_2,*}$ when $P_\oned\gg 1$, and when we assume a perfect timing,  $T=\pi/\sqrt{(k+1) \Gamma_\oned \Gamma^*}$ such that no population remains in the dark state, $|\phi_d(T)|^2 = 0$. The only contribution remaining is the one of the superradiant state
$\ket{\phi_s(T)} = \sqrt{\frac{k+1}{N_\bb + k+1}} e^{-(\Gamma^*+(N_\bb + k + 1) \Gamma_{\oned})T/2}$, which leads to
\begin{align}
p_{\aa_2,*}\lesssim \frac{k+1}{(N_\bb + k+1)^2 P_\oned} e^{-\pi (N_\bb + k + 1) / \sqrt{(k+1)P_{\oned}}}\,, \\
p_{\bb_2,*}\lesssim \frac{N_\bb}{(N_\bb + k+1)^2 P_\oned} e^{-\pi (N_\bb + k + 1) / \sqrt{(k+1)P_{\oned}}} \,,
\end{align}
which are negligible compared to $p_{\aa,\bb_1,*}$ for sufficiently large $N_\bb$.

\section{Quantum Zeno dynamics in protocol 3}\label{appendix: zeno2}

The main difference with respect to the other protocols is that is done in a single step. We start with  $\ket{\psi_1} = \ket{e}_\rs \ket{\phi_m^{\{s_n\}}}_\rt \ket{s}_\rd$ and consider that only the classical field $\Omega_{ge}^\rd\equiv \Omega_\rd \neq 0$, such that the dynamics are governed by the effective Hamiltonian
\begin{align} \label{eqSM:Hameff}
H_\mathrm{eff}
	=& \frac{\Omega^\rd}{2}\big(\sigma_{ge}^\dtc + \sigma_{eg}^\dtc \big) 
		-\ii \frac{\Gamma_{\oned}^\rg}{2}\left(\sigma_{eg}^\src + S_{eg}^\trg\right) \left(\sigma_{ge}^\src+S_{ge}^\trg \right) \nonumber \\ 
	&-\ii \frac{\Gamma_{\oned}^\rs}{2} \left(S_{es}^\trg + \sigma_{es}^\dtc \right) 
		\left(S_{se}^\trg + \sigma_{se}^\dtc \right)
	-\ii \frac{\Gamma^*}{2}\sum_{n}\sigma^{n}_{ee} \,,
\end{align}
which couples the initial state $\ket{\psi_1}$ to the state with the population in the excited state in the target ensemble,
$\ket{\psi_2} \propto \ket{g}_\rs S_{eg}\ket{\phi_m^{\{s_n\}}}_\rt \ket{s}_\rd$, in the detector ensemble,
$\ket{\psi_3} \propto \ket{g}_\rs S_{sg}\ket{\phi_m^{\{s_n\}}}_\rt \ket{e}_\rd$ and the goal state,
$\ket{\psi_4} = \ket{g}_\rs \ket{\phi_{m+1}^{\{s_n'\}}}_\rt \ket{g}_\rd$, where $\{s_n'\}$ differs from $\{s_n\}$ only by an additional excitation in $s$. In particular $\ket{\phi_{m+1}^{\{s_n'\}}} \propto S_{sg}\ket{\phi_m^{\{s_n\}}}$.
In this basis the Hamiltonian can be written as
\begin{widetext}
\begin{equation*}
	H_{\mathrm{eff}} = 
	\frac{1}{2} \left( \begin{array}{cccc} 
	-i (\Gamma_\oned^\rg+\Gamma^*) & -i \sqrt{N_m} \Gamma_\oned^\rg & 0  & 0 \\
	-i \sqrt{N_m}\Gamma_\oned^\rg  & - i (N_m \Gamma_\oned^\rg+\Gamma_\oned^\rs+\Gamma^*) & -i \Gamma_\oned^\rs  & 0 \\
	0 & -i \Gamma_\oned^\rs & -i (\Gamma_\oned^\rs+\Gamma^*) & \Omega^\rd \\
	0 & 0 & \Omega^\rd & 0
	\end{array} \right)\,,
	\label{eq:Ham1step}
\end{equation*}
\end{widetext}

The goal is to find 
$\ket{\psi(t)}=e^{-i H_\mathrm{eff} t}\ket{\psi_1}=\sum_{j} c_j(t)\ket{\psi_j}$,
which can easily be done numerically as it acts on a four dimensional Hilbert space. However, in order to gain more insight it is useful to write the Hamiltonian in a basis of sub- and superradiant states. Interestingly, the system always contains a dark state 
$\ket{\psi_\rd}=\frac{1}{\sqrt{N_m+2}} \left( \sqrt{N_m} \ket{\psi_1}-\ket{\psi_2}+\ket{\psi_3}\right)$,
which is dark for all values of $\Gamma_\oned^\rg$ and $\Gamma_\oned^\rs$. However, the two orthogonal excited states depend on the ratio $\Gamma_\oned^\rs / \Gamma_\oned^\rg$. After a careful study, we found the optimal choice for the maximum transfer to $\ket{\psi_4}$ was to choose  $\Gamma_\oned^\rs / \Gamma_\oned^\rg = \frac{N_m+1}{2}$. The reason is that in that case the two decay channels to $\ket{g}$ and $\ket{s}$ show the same "superradiant" decay $(N_m+1)\Gamma_\oned^\rg$.  From now on, we use that ratio and $\Gamma_{\oned}
^g=\Gamma_{\oned}$. For this ratio, the adiabatic elimination of the superradiant states gives rise to an effective non-hermitian Hamiltonian between the dark state and the goal state, which reads
\begin{equation}\label{eqSM:eff1step}
H_{\mathrm{eff}} 
	\approx \left(
	\begin{array}{cc}
 		-\frac{i \Gamma^*}{2} &  \frac{\Omega_\rd}{2\sqrt{N_m}} \\
   		\frac{\Omega_\rd}{2\sqrt{N_m}} & -i\frac{3 \Omega_\rd^2 }{2N_m\Gamma_\oned}\\
	\end{array}\right)\,.
\end{equation}
where we have used $N_m\gg 1$ to simplify the expressions. As we know from the previous Sections, one should choose an $\Omega_\rd$ such that the contribution of $\Gamma^*$ and the effective losses induced by populating the superradiant states, i.e., $\frac{3(\Omega^\rd)^2 }{2N_m\Gamma_\oned}$, are equal. Thus, by choosing $\Omega_\rd = \sqrt{N_m \Gamma_\oned \Gamma^* / 3}$, it is easy to find that:
\begin{align}\label{eqSM:a}
|c_\rd(t)|^2 
	&\approx \frac{N_m}{N_m+2} e^{-\Gamma^*t} \cos^2(\frac{\sqrt{\Gamma_\oned\Gamma^*}t}{2\sqrt{3}})\,, \nonumber \\
|c_4(t)|^2
	&\approx \frac{N_m}{N_m+2} e^{-\Gamma^*t} \sin^2(\frac{\sqrt{\Gamma_\oned\Gamma^*}t}{2\sqrt{3}})\,.
\end{align}

Therefore, if $P_\oned\gg1$ and 
$T_\mathrm{opt}^\bb =\frac{\pi\sqrt{3}}{\sqrt{\Gamma_{\oned}\Gamma^*}}$ 
then the probability of heralding reads:
\begin{align} \label{eqSM:pbig}
p_{m \rightarrow m+1}
	= |c_4(T_\mathrm{opt}^\bb)|^2
	\approx \frac{N_m}{N_m+2} e^{-\sqrt{3}\pi/\sqrt{P_\oned}}\,,
\end{align}

The problematic quantum jump in this case is related to the probability of having an excitation $\ket{e}$ in the \emph{target} ensemble, i.e., coming from the population of state $\ket{\psi2}$ that reads
\begin{align}\label{eqSM:pbad11}
p_{*}
	=&p_{1,*} + p_{2,*}
	=\Gamma^*\int_0^{T_\mathrm{opt}^\bb} \rd t |c_2(t)|^2 \\
	& + \Gamma^*\int_{0}^\infty \rd t \sum_{j = \pm} |\langle \psi_{\rs,j} | \psi_2 \rangle|^2  e^{-\left( (N_m+1)\Gamma_\oned + \Gamma^*\right) t}\,, \nonumber
 \end{align}
where the first part corresponds to the interval of time $(0,T_\mathrm{opt}^\bb)$ where $\Omega_\rd$ is switched on, and the second part from populating the two superradiant states. The contributions of the superradiant/subradiant states to $\ket{\psi_2}$ can be obtained in the asymptotic limit $N_m\gg 1$, where:
\begin{align}\label{eqSM:overlap}
|\braket{\psi_\rd}{\psi_2}
	&\sim \frac{1}{\sqrt{N_m}}\,,\\
|\braket{\psi_{\rs,\pm }}{\psi_2}|
	&\sim \frac{\sqrt{2\pm \sqrt{2}}}{2}\sim O(1)\,.
\end{align}

Using that information we can estimate the contribution of the dark state within the time interval $(0,T_\mathrm{opt}^\bb)$.
\begin{align}\label{eq:pbad11}
p_{1,*}
	&\approx \frac{\Gamma^*}{2 (N_m+2)} \int_0^{T_\mathrm{opt}^\bb} \rd t\ e^{-\Gamma^* t}
	\lesssim \frac{1-e^{-\Gamma^* T_\mathrm{opt}^\bb}}{2(N_m+2)} \\
	&\approx \frac{\pi \sqrt{3}}{2 N_m\sqrt{P_\oned}}\,,\, \mathrm{for}\, P_\oned, N_m\gg 1 \,.
 \end{align}
The contribution of $p_{2,*}$ can always be shown to be smaller order when $P_\oned\gg 1$. Therefore, the overall probability of emitting a quantum jump in the target ensemble for this protocol 
\begin{align}
p_{*}
&\approx \frac{\pi\sqrt{3}}{2N_m\sqrt{P_\oned}}\,,\,\mathrm{if}\,\,\,, P_\oned\gg 1\,.
\end{align}
The repumping process to correct the collective quantum jump errors can be done as well in a single step by pumping collectively the atoms from $s\rightarrow g$ , which moves $S_{sg}\ket{\Phi_m}\rightarrow S_{eg}\ket{\Phi_m}\rightarrow\ket{\Phi_m}$.  Because this process is done through a collective photon the probability of emitting a free space photon is given by $p_{\mathrm{pump},*} \propto \frac{1}{NP_\oned}$.

\subsection{Fidelity of the protocol.}

The analysis of the fidelity is very similar to the one performed for the protocols 1 and 2. In this case, if we detect an excitation in $g$ in the detector atom, no error can appear. Therefore, the only contribution to the infidelity is the error coming from free space photons in each attempt after $1/p$ repetitions, which finally scale:
\begin{equation}
I_{m\rightarrow m+1}
	\approx \frac{1}{p} (p_{*}+p_{\mathrm{pump},*}) \frac{m}{N_m}
	\propto \frac{m}{N_m^2 \sqrt{P_\oned}}\,,
\end{equation}
for systems with $P_\oned,N_m\gg 1$. If we consider as a reference the largest Purcell factor (the one associated to $\Gamma_{\oned}^\rs \approx N_m \Gamma_\oned / 2$), then, the scaling of probabilities and fidelities should read equivalently
\begin{align}\label{eq:fidsinglestep}
p_{m \rightarrow m+1}
	&\approx \frac{N_m}{N_m+2} e^{-\sqrt{3}\pi \sqrt{N_m/2}/\sqrt{P_\oned^\rs}}\,\nonumber \\
I_{m\rightarrow m+1}
	&\approx \frac{m \sqrt{2}}{N_m^{3/2} \sqrt{P_\oned^\rs}}\,,
\end{align}
where we replaced 
$P_\oned =\frac{ \Gamma_\oned^\rg}{\Gamma^*} \approx \frac{2}{N_m} \frac{\Gamma_\oned^\rs}{\Gamma^*} = \frac{2}{N_m} P_\oned^\rs$.

\end{document}